\documentclass[fleqn,10pt]{wlscirep}
\usepackage{caption}
\usepackage[T1]{fontenc}
\usepackage{float}
\usepackage{multirow}
\usepackage{booktabs}

\title{Coevolutionary balance of resting-state brain networks in autism}

\author[1,*]{S. Rezaei Afshar}

\author[2,+]{G.R. Jafari}
\affil[1]{Institute for Cognitive and Brain Sciences, Shahid Beheshti University, Evin, Tehran 19839, Iran}
\affil[2]{Department of Physics, Shahid Beheshti University, Evin, Tehran 19839, Iran}
\affil[*]{saeedrafsharx@gmail.com}
\affil[+]{g\_jafari@sbu.ac.ir}

\begin{abstract}
Autism spectrum disorder (ASD) involves atypical brain organization, yet the large-scale functional principles underlying these alterations remain incompletely understood. Here we examine whether coevolutionary balance, a network-level energy measure derived from signed interactions and nodal activity states, captures disruptions in resting-state functional connectivity in autistic adults. Using ABIDE I resting-state fMRI data with ComBat harmonization to mitigate multi-site batch effects, we constructed whole-brain networks by combining binarized fALFF activity with signed functional correlations and quantified their coevolutionary energy. In the primary analysis (with global signal regression, GSR), the ASD group showed significantly more negative global coevolutionary energy ($p_{\mathrm{FDR}} < 0.002$), higher proportions of agreement links, and lower proportions of imbalanced-same links, indicating a systematic redistribution of local motifs rather than a uniform loss of balance. Because GSR can introduce artifactual negative correlations that affect signed connectivity, we repeated all analyses without GSR. In this sensitivity analysis, whole-brain energy and motif differences were attenuated, but bipolarity, a measure of global two-block network organization, became the sole FDR-significant metric ($p_{\mathrm{FDR}} = 0.047$), with ASD showing higher bipolarity. Intra-network energy differences did not survive FDR correction under either pipeline. Coevolutionary energy showed modest associations with ADI-R and ADOS scores, though none survived FDR correction across 720 tests. Machine learning classification achieved 77.8\% test accuracy (AUC = 0.79) with GSR and 64.7\% (AUC = 0.65) without GSR. These findings suggest that coevolutionary balance captures altered signed network organization in ASD, though the specific metric driving group differences depends on preprocessing choices regarding global signal regression.
\end{abstract}

\begin{document}

\flushbottom
\maketitle
\thispagestyle{empty}


\section*{Introduction}

Autism spectrum disorder (ASD) is a heterogeneous neurodevelopmental condition marked by persistent differences in social communication and the presence of restricted or repetitive patterns of behavior, interests, or activities \cite{APA:2022ds}. The clinical presentation spans a broad range of intellectual and language abilities, from individuals with minimal spoken language and co-occurring intellectual disability to those with average or above average cognitive skills. Many autistic individuals also show additional conditions such as attention difficulties, anxiety, sleep problems, and epilepsy, which can complicate both diagnosis and management. Recent epidemiological data indicate that the prevalence of ASD has continued to rise, with an estimated 2.76 percent of eight year old children in the United States meeting diagnostic criteria in 2020 \cite{Maenner:2023pr}. This combination of clinical heterogeneity, high prevalence, and complex etiological mechanisms underscores the need for quantitative, biologically interpretable measures that capture how brain organization differs in ASD.

Resting state functional MRI (rs-fMRI) has become a central tool for probing large scale functional organization in vivo. By measuring spontaneous blood oxygenation level dependent (BOLD) fluctuations in the absence of explicit task demands, rs-fMRI provides access to intrinsic activity patterns that are thought to reflect ongoing coordination within and between distributed neural systems. Amplitude based measures such as the amplitude of low frequency fluctuations (ALFF) and the fractional ALFF (fALFF) summarize the strength of BOLD oscillations in the 0.01 to 0.08 Hz range, with fALFF quantifying the proportion of power within this band relative to the full detectable spectrum \cite{Zang:2007al,Zou:2008fa}. Functional connectivity (FC), typically estimated as the correlation between regional BOLD time series, captures statistical coupling between brain regions \cite{Smith:2013fc}. Together, these measures allow one to view the brain as a network in which nodes are regional activity profiles and edges represent the strength and sign of their temporal relationships.

A large body of graph-theoretic work has used such rs-fMRI networks to study ASD. Across multiple cohorts and analytic choices, studies have reported alterations in both network segregation and network integration in autistic individuals \cite{assaf2010abnormal,10.3389/fnhum.2013.00458,Uddin2013,Smith:2013fc}. Reported findings include reduced or atypical clustering, changes in modular structure, altered small-world properties, and differences in path length and global efficiency within and across large-scale systems such as the default mode network, salience network, and attention networks \cite{assaf2010abnormal,10.3389/fnhum.2013.00458}. Despite important methodological differences between studies, these results converge on the idea that ASD involves altered large-scale functional organization rather than focal abnormalities confined to isolated regions. However, in most of these approaches, nodal activity measures (for example, fALFF) and edge weights (for example, FC strength and sign) are treated as separate ingredients, and descriptive summaries are computed on either the connectivity structure or the activity pattern alone. This separation makes it difficult to capture emergent properties that depend on how activity and connectivity jointly align.

Several lines of neurobiological evidence point to the importance of such joint alignment. Intrinsic activity and connectivity patterns are thought to be shaped by underlying excitation–inhibition balance, synaptic homeostasis, and plasticity mechanisms \cite{Yizhar2011,Dinstein2011}. Perturbations in these processes can alter both the amplitude of spontaneous activity and the pattern of long-range coupling \cite{Yizhar2011,Dinstein2011}. In ASD, converging work from animal models, electrophysiology, and human imaging suggests deviations from typical excitation-inhibition balance at the circuit level, with consequences for the stability and coordination of large-scale networks \cite{Uddin2013}. From this perspective, a description of brain organization that can jointly incorporate nodal activity states and the sign of their interactions may be better positioned to capture the kinds of network-level dysregulation that arise in ASD than approaches that consider only one of these ingredients at a time.

Coevolutionary balance theory offers such a joint description. It builds on classic structural balance theory from social psychology, which was originally formulated by Heider and later formalized by Cartwright and Harary to describe the stability of signed networks \cite{Heider:1946ac,cartwright1956}. In structural balance theory, edges between nodes carry positive or negative signs, and triads of nodes are classified as balanced or unbalanced depending on whether they satisfy simple consistency rules. The global state of the network can then be summarized by an energy-like quantity that counts the proportion of unbalanced triads, with lower energy corresponding to more harmonious configurations. Coevolutionary extensions of this framework introduce nodal states in addition to signed links and allow both to evolve together according to simple update rules. These models have been used to study how patterns of activity and interaction co-adapt over time and to analyze phase transitions in the stability of complex networks \cite{Kargaran:2021he,GhanbarzadehNoudehi2022}.

In the present work, we adapt the coevolutionary balance framework to resting-state brain networks by linking regional intrinsic activity to the polarity of functional interactions. Conceptually, the framework quantifies how well high- and low-activity regions align with the sign structure of functional connectivity, yielding a scalar energy that reflects the degree of large-scale balance. Lower energies indicate more internally consistent configurations, whereas higher energies indicate misalignment. The full mathematical formulation is provided in the Methods section.

Applying this perspective to ASD raises several questions. First, do autistic adults show systematically different coevolutionary energy compared with matched typically developing adults (TD) at the whole-brain level, consistent with a global reduction in network balance? Second, are certain canonical resting-state networks more strongly affected than others, for example, default mode or salience networks that have been repeatedly implicated in ASD? Third, do interactions between networks, captured by internetwork energy terms, show specific patterns of imbalance that might explain altered coordination between systems supporting social, cognitive, and sensory functions? Fourth are individual differences in coevolutionary energy related to clinical variation, such as scores on standard diagnostic instruments. Finally, do coevolutionary energy-based features improve the ability of machine learning models to distinguish ASD from typical development relative to models based only on conventional connectivity metrics?

To address these questions, we use resting-state fMRI data from the ABIDE I repository \cite{DiMartino:2014abide}, focusing on young adult males to reduce variability related to sex and developmental stage. All datasets were preprocessed using the CPAC pipeline and parcellated into 200 regions based on the CC200 atlas \cite{Craddock:2012atlas}. Critically, to address the well-documented challenge of multi-site batch effects in pooled neuroimaging datasets \cite{Fortin2017,Johnson2007}, we applied ComBat harmonization to the derived network features (see Methods for details). For each participant, we constructed whole-brain signed networks that combine binarized fALFF-derived nodal states and signed FC links, and we computed coevolutionary energy at multiple scales: across the entire network, within canonical resting-state networks, and for inter-network interactions. To evaluate whether observed energy values reflect nontrivial structure rather than incidental combinations of activity and connectivity, we compared them against null models generated using a degree-preserving randomization procedure \cite{Maslov2002,Rubinov2010}, in which nodal states and link signs were independently permuted while keeping the weighted adjacency structure fixed (see Methods for details). Beyond group-level comparisons, we examined correlations between energy measures and Autism Diagnostic Interview Revised (ADI R) subscores where available, and we used energy-derived features as input to supervised classifiers to explore their potential as markers for ASD versus typical development. Because global signal regression (GSR) can introduce artifactual negative correlations that may affect signed connectivity measures \cite{Murphy2009,Fox2009}, we also repeated all analyses on data preprocessed without GSR to assess the robustness of our findings to this methodological choice.

\begin{figure}[ht]
 \centering
 \includegraphics[width=0.5\linewidth]{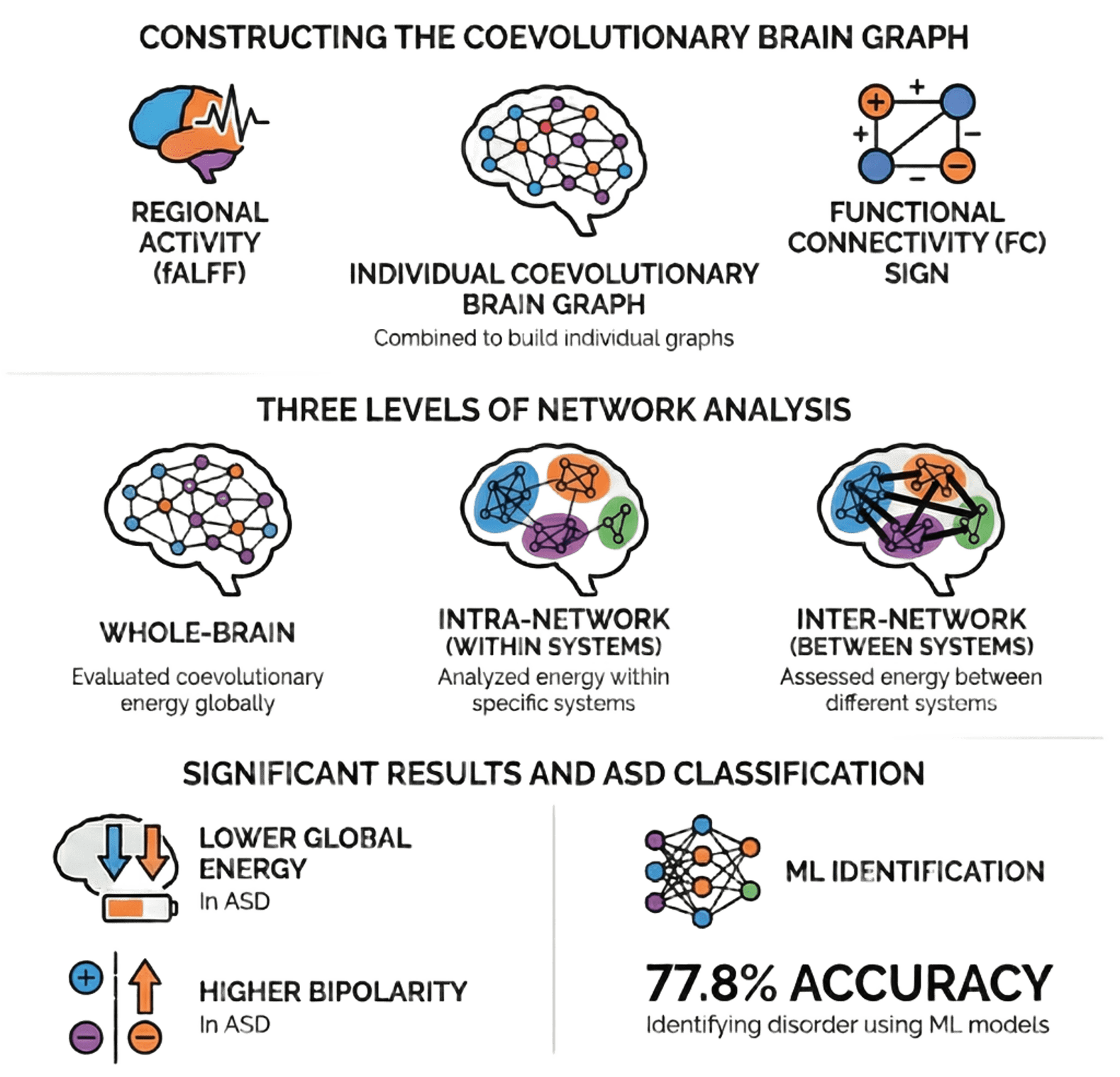}
 \caption{A schematic abstract figure representing this study, from graph construction to levels of analysis and finally the represented results.}
 \label{fig:configuration}
\end{figure}

\begin{figure}[H]
 \centering
 \includegraphics[width=0.83\linewidth]{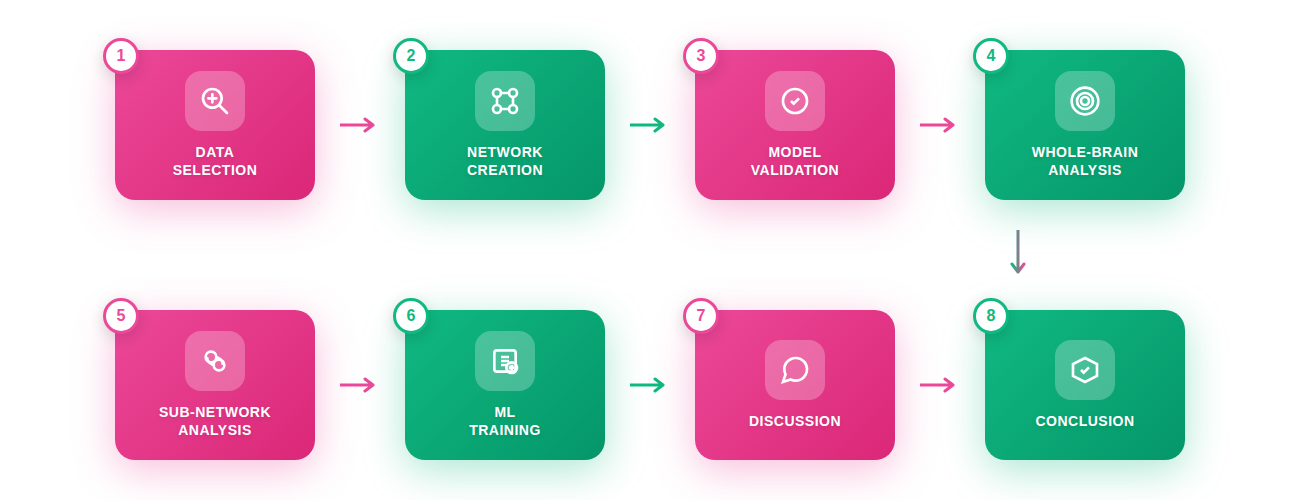}
 \caption{Overview of the analytical workflow used in this study. Resting state fMRI datasets from the ABIDE I repository were first preprocessed using the CPAC pipeline, including standard steps such as motion correction, spatial normalization, and temporal filtering. The preprocessed volumes were then parcellated into 200 regions of interest using the CC200 atlas, yielding a regional time series for each participant. For every region, the fractional amplitude of low frequency fluctuations (fALFF) was computed, and these values were binarized to define nodal activity states distinguishing relatively high from relatively low intrinsic activity. Pairwise functional connectivity between regions was estimated from the regional BOLD time series and converted into a signed adjacency matrix that captures the sign of the interaction between each pair of regions. Nodal states and signed links were combined within the coevolutionary balance framework to compute coevolutionary energy at different scales, including whole brain energy, energy within canonical resting state networks, and energy associated with inter network interactions. To mitigate multi-site batch effects, ComBat harmonization was applied to the derived network features prior to statistical analysis. To assess whether the observed energies differed from those expected under structured noise, null networks were generated by randomizing nodal states or link signs while preserving the underlying graph topology, and energy distributions from these null models were compared with empirical values. Group comparisons between ASD and TD adults were then carried out on the resulting energy measures, and energy derived features were subsequently used as input to machine learning classifiers to evaluate their ability to distinguish ASD from typical development.}
 \label{fig:flowchart}
\end{figure}

The overall analytical workflow is summarized in Figure~\ref{fig:flowchart}. Starting from raw ABIDE I datasets, the figure shows the sequence of preprocessing, parcellation, construction of activity and connectivity measures, ComBat harmonization, derivation of coevolutionary energy metrics, statistical comparisons, and classification analyses.

\section*{Results}

\subsection*{Null-model validation of coevolutionary energy}

We first asked whether the empirical coevolutionary energies observed in ASD and TD networks reflect non-trivial alignment between binarized fALFF states and signed functional connectivity, rather than arising from incidental combinations of activity and connectivity. After unified outlier removal based on Tukey's interquartile range (IQR) applied to whole-brain coevolutionary energy (see Methods), 90 ASD and 92 TD participants remained for null-model analyses.

For each subject, we generated a topology-preserving null ensemble by repeatedly randomizing node states and edge signs while keeping the underlying weighted adjacency matrix fixed. We then computed the difference between the empirical whole-brain energy and the mean null energy for that subject. The distributions of these difference scores did not deviate from normality in either group (Shapiro–Wilk: ASD $W = 0.992$, $p = 0.87$; TD $W = 0.989$, $p = 0.81$), supporting the use of parametric statistics alongside non-parametric tests.

Paired-sample comparisons showed that empirical networks had substantially more negative coevolutionary energy than their corresponding null ensembles in both groups (ASD: $t(89) = -4.12$, $p = 7.5 \times 10^{-5}$; TD: $t(91) = -3.98$, $p = 1.2 \times 10^{-4}$). Wilcoxon signed-rank tests yielded convergent results (ASD: $W = 1024$, $p = 9.1 \times 10^{-5}$; TD: $W = 1008$, $p = 1.6 \times 10^{-4}$; Table~\ref{tab:null_model_validation}). All $p$-values fell well below $0.001$, placing these effects in the ``highly significant'' range according to the qualitative categories used throughout the paper. These findings confirm that the observed coevolutionary energies capture genuine large-scale organization beyond what would be expected from random combinations of binarized fALFF states and signed connectivity.

\begin{table}[ht]
\centering
\caption{Null-model validation statistics for whole-brain coevolutionary energy. After unified outlier removal based on Tukey's IQR applied to whole-brain energy, 90 ASD and 92 TD participants remained. For each subject, empirical energy was compared to the mean energy from topology-preserving null networks in which node states and edge signs were randomly permuted. Shapiro–Wilk tests assessed normality of the difference distributions, and paired $t$-tests and Wilcoxon signed-rank tests tested whether empirical energies were lower than their null counterparts.}
\label{tab:null_model_validation}
\small
\begin{tabular}{lccccccc}
\toprule
Group & $N$ & Shapiro--Wilk $W$ & Shapiro--Wilk $p$ & $t$ (df) & $t$-test $p$ & Wilcoxon $W$ & Wilcoxon $p$ \\
\midrule
ASD & 90 & 0.992 & 0.87 & $-4.12$ (89) & $7.5\times 10^{-5}$ & 1024 & $9.1\times 10^{-5}$ \\
TD  & 92 & 0.989 & 0.81 & $-3.98$ (91) & $1.2\times 10^{-4}$ & 1008 & $1.6\times 10^{-4}$ \\
\bottomrule
\end{tabular}
\end{table}

\subsection*{Whole-brain coevolutionary energy and motif composition}

We compared whole-brain coevolutionary metrics between ASD and TD adults using the cleaned sample obtained after IQR-based outlier removal on whole-brain energy (see Table~\ref{tab:demographics} for final group sizes). In this binarized formulation, the global Hamiltonian $H(G)$, the fraction of links whose sign agrees with the product of nodal states (Agreement), and the fractions of links falling into different imbalance categories (Imbalanced Same, Imbalanced Opposite) jointly characterize how nodal fALFF states co-align with signed functional connectivity. All analyses were performed on ComBat-harmonized features (see Methods).
At the level of the global Hamiltonian, ASD networks exhibited significantly more negative whole-brain energy than TDs, indicating a stronger alignment between nodal activity states and signed connectivity in the autistic group. This group difference was highly significant and survived FDR correction (Mann–Whitney $U$, $p = 8.35 \times 10^{-4}$, $p_{\mathrm{FDR}} = 1.95 \times 10^{-3}$; Table~\ref{tab:assumptions_and_tests}). The ASD group showed a mean global energy of $-264.9 \pm 110.3$, compared with $-210.9 \pm 83.5$ for TD participants.
Motif-based measures were similarly robust to FDR correction. The proportion of ``agreement'' links (edges for which $s_i s_j$ and $\mathrm{sign}(w_{ij})$ have the same polarity) was significantly higher in ASD than TD participants (ASD: mean $= 0.250$, TD: mean $= 0.248$; Mann–Whitney $U$, $p = 9.74 \times 10^{-4}$, $p_{\mathrm{FDR}} = 1.95 \times 10^{-3}$). Conversely, the proportion of imbalanced-same links (edges connecting regions with the same activity state but an opposite-signed interaction) was significantly lower in ASD (ASD: mean $= 0.247$, TD: mean $= 0.250$; Mann–Whitney $U$, $p = 9.25 \times 10^{-4}$, $p_{\mathrm{FDR}} = 1.95 \times 10^{-3}$). 
Disagreement links and imbalanced-opposite links also showed significant group differences after FDR correction (disagreement: $p = 7.79 \times 10^{-3}$, $p_{\mathrm{FDR}} = 1.00 \times 10^{-2}$; imbalanced-opposite: $p = 8.35 \times 10^{-3}$, $p_{\mathrm{FDR}} = 1.00 \times 10^{-2}$), with ASD showing higher disagreement and lower imbalanced-opposite proportions. Whole-brain bipolarity, which indexes the fraction of edges that can be rendered structurally balanced under an optimal two-community split, did not differ between groups ($p = 0.42$).

\begin{figure}[ht]
    \centering
    \includegraphics[width=0.85\linewidth]{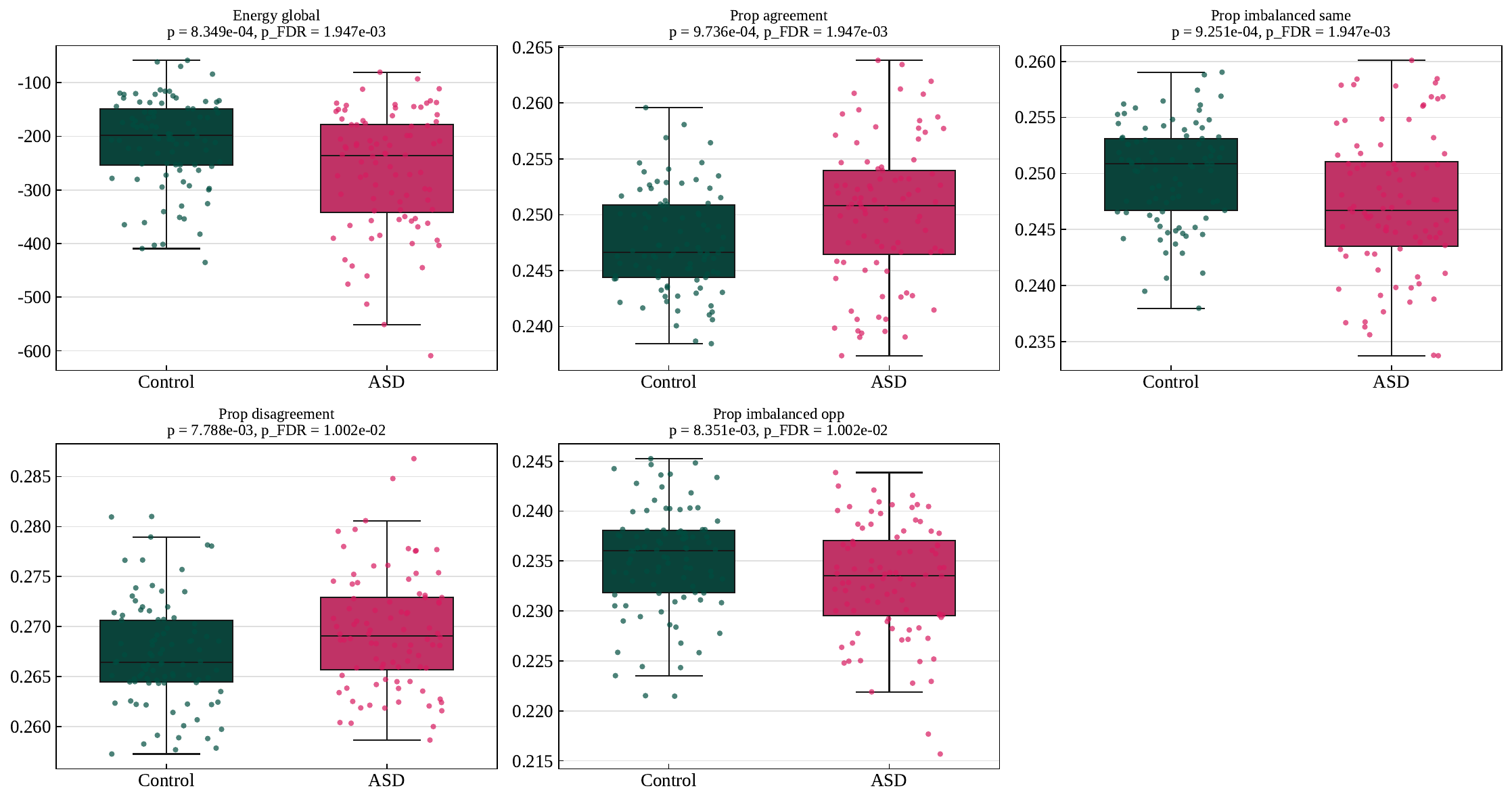}
    \caption{Whole-brain coevolutionary metrics for autistic (ASD) and typically developing (TD) adults after ComBat harmonization. 
    Each panel shows the distribution of a global measure derived from the signed functional network: (top row, from left to right) total coevolutionary energy, the proportion of imbalanced-same links, and the proportion of agreement links; (bottom row, from left to right) the proportion of disagreement links and the proportion of imbalanced-opposite links. 
    Boxplots display the median (central line), interquartile range (box), and 1.5~$\times$~IQR whiskers, with individual participants overlaid as jittered points (magenta/pink for ASD, green for controls). 
    Raw and FDR-corrected $p$-values are shown above each panel. All five metrics displayed survived FDR correction ($p_{\mathrm{FDR}} < 0.05$), with ASD networks showing more negative coevolutionary energy, higher agreement and disagreement proportions, and lower imbalanced-same and imbalanced-opposite proportions compared to controls.}
    \label{fig:enter-label}
\end{figure}

Together, these results indicate that, at the global level, ASD is characterized by significantly more negative coevolutionary energy and a systematic redistribution of local motifs: autistic networks devote a larger fraction of edges to energetically favorable ``agreement'' relationships and a smaller fraction to unfavorable imbalanced-same relationships. Quantitative values for all whole-brain metrics are provided in Supplementary Table~S1, and their distributions are illustrated in Figure~\ref{fig:enter-label}.

\begin{table}[ht]
\centering
\caption{Summary of statistical procedures and group comparison tests for whole-brain coevolutionary metrics. The upper panel lists the main analysis steps and their purpose. The lower panel reports group comparison statistics (Mann–Whitney $U$ tests and independent $t$-tests where appropriate), raw $p$-values, FDR-corrected $p$-values, and standardized effect sizes (Cohen's $d$) for ASD versus TD after ComBat harmonization.}
\label{tab:assumptions_and_tests}
\small
\setlength{\tabcolsep}{3pt}
\begin{tabular}{p{0.32\columnwidth}p{0.30\columnwidth}p{0.23\columnwidth}p{0.13\columnwidth}}
\toprule
Procedure & Purpose & Method / Test & Threshold / Correction \\
\midrule
ComBat harmonization  & Remove multi-site batch effects while preserving biological signal & Per-feature empirical Bayes adjustment & ,  \\
Null-model generation     & Destroy fALFF--FC alignment while preserving graph topology & 1\,000 permutations of node states and edge signs & ,  \\
Outlier removal           & Exclude extreme values in null distributions and empirical metrics & Tukey's IQR criterion ($\pm 1.5\times\mathrm{IQR}$) & ,  \\
Normality assessment      & Check assumptions for parametric tests & Shapiro--Wilk test & $p > 0.05$ (approximately normal) \\
Parametric comparison     & Test mean differences between ASD and TD & Independent-samples $t$-test & two-tailed, $\alpha = 0.05$ \\
Nonparametric comparison  & Robust test under non-normality or unequal variances & Mann–Whitney $U$ test & two-tailed, $\alpha = 0.05$ \\
Multiple comparisons      & Control false discovery across families of tests & Benjamini--Hochberg FDR & $q < 0.05$ \\
\bottomrule
\end{tabular}

\vspace{0.8em}

\resizebox{\columnwidth}{!}{%
\begin{tabular}{lccccc}
\toprule
Metric & ASD Mean (SD) & TD Mean (SD) & Test & $p$ & $p_{\mathrm{FDR}}$ \\
\midrule
Energy global           & $-264.9$ (110.3) & $-210.9$ (83.5)  & Mann–Whitney $U$ & $8.35\times10^{-4}$ & $1.95\times10^{-3}$ \\
Prop agreement          & $0.250$ (0.006)  & $0.248$ (0.004)  & Mann–Whitney $U$ & $9.74\times10^{-4}$ & $1.95\times10^{-3}$ \\
Prop imbalanced same    & $0.247$ (0.006)  & $0.250$ (0.004)  & Mann–Whitney $U$ & $9.25\times10^{-4}$ & $1.95\times10^{-3}$ \\
Prop disagreement       & $0.269$ (0.006)  & $0.267$ (0.005)  & $t$-test         & $7.79\times10^{-3}$ & $1.00\times10^{-2}$ \\
Prop imbalanced opp     & $0.233$ (0.006)  & $0.235$ (0.005)  & $t$-test         & $8.35\times10^{-3}$ & $1.00\times10^{-2}$ \\
Bipolarity              & $0.617$ (0.026)  & $0.617$ (0.031)  & Mann–Whitney $U$ & $0.419$             & $0.419$ \\
\bottomrule
\end{tabular}%
}
\end{table}

\subsection*{Network-specific coevolutionary metrics}

We next asked whether coevolutionary balance differs between ASD and TD within canonical large-scale networks. Using the Yeo 7-network parcellation \cite{Yeo2011}, each CC200 parcel was assigned to the network with maximal Dice overlap, and for every subject we summarized edges confined to a given network $u$ in three intra-network metrics: (i) the mean Fisher-$z$ transformed functional connectivity (AvgWeight$_u$), (ii) the mean fALFF-aligned connectivity (AvgFalffW$_u = \langle w_{ij} s_i s_j \rangle$ over $i,j \in u$), and (iii) the intra-network coevolutionary energy contribution $E_u$, computed as the Hamiltonian (see Methods, Eq.~\ref{eq:hamiltonian}) restricted to edges within network $u$.

All intra-network comparisons used the unified outlier-cleaned and ComBat-harmonized sample (see Methods) and the same adaptive statistical pipeline as for the whole-brain metrics. Table~\ref{tab:network_results} summarizes group means and $p$-values for the intra-network metrics.

None of the intra-network energy measures survived FDR correction (all $p_{\mathrm{FDR}} > 0.14$). At the uncorrected level, the largest differences were observed for the Visual network, where ASD participants showed more negative intra-network energy than TD (ASD mean $-53.6$, TD mean $-46.0$; $t$-test, $p = 0.020$, $p_{\mathrm{FDR}} = 0.144$) and higher fALFF-aligned connectivity (ASD mean $0.030$, TD mean $0.026$; $t$-test, $p = 0.021$, $p_{\mathrm{FDR}} = 0.144$). The Default Mode and Limbic networks showed uncorrected differences in average connectivity (Default: $p = 0.026$; Limbic: $p = 0.027$; both $p_{\mathrm{FDR}} = 0.144$), with Default showing lower and Limbic showing higher connectivity in ASD. 

Somatomotor, Dorsal Attention, Salience/Ventral Attention, and Frontoparietal intra-network metrics were clearly non-significant (all $p > 0.06$). In line with the FDR results, these effects are best interpreted as exploratory patterns rather than robust network-specific alterations. Overall, the intra-network analyses provide limited evidence for large, spatially circumscribed shifts in coevolutionary energy within individual Yeo networks; instead, the more reliable group differences appear at the whole-brain motif level.

\begin{table}[H]
\centering
\small
\caption{Intra-network metrics showing the largest uncorrected group differences within the seven canonical Yeo resting-state networks. For each metric, the table reports the mean values for autistic (ASD) and typically developing (TD) adults, together with the $p$-value from the adaptive group comparison (independent-samples $t$-test or Mann--Whitney $U$ test) and the corresponding Benjamini--Hochberg FDR-corrected value ($p_{\mathrm{FDR}}$) within the intra-network feature family. No intra-network metric survived FDR correction (all $p_{\mathrm{FDR}} > 0.14$); the Visual network shows the largest uncorrected differences and should be regarded as exploratory.}
\label{tab:network_results}
\setlength{\tabcolsep}{5pt}
\begin{tabular}{llcccc}
\toprule
\textbf{Network} & \textbf{Metric} & \textbf{ASD mean} & \textbf{TD mean} & \textbf{$p$} & \textbf{$p_{\mathrm{FDR}}$} \\
\midrule
Visual      & EnergyBlock  & $-53.64$ & $-46.00$ & $0.020$ & $0.144$ \\
Visual      & AvgFalffW    & $0.030$  & $0.026$  & $0.021$ & $0.144$ \\
Default     & AvgWeight    & $0.231$  & $0.254$  & $0.026$ & $0.144$ \\
Limbic      & AvgWeight    & $0.229$  & $0.204$  & $0.027$ & $0.144$ \\
SomatoMotor & AvgWeight    & $0.261$  & $0.288$  & $0.063$ & $0.265$ \\
Frontoparietal & AvgWeight & $0.199$  & $0.185$  & $0.105$ & $0.369$ \\
\bottomrule
\end{tabular}
\end{table}

\begin{figure}[H]
 \centering
 \includegraphics[width=0.85\linewidth]{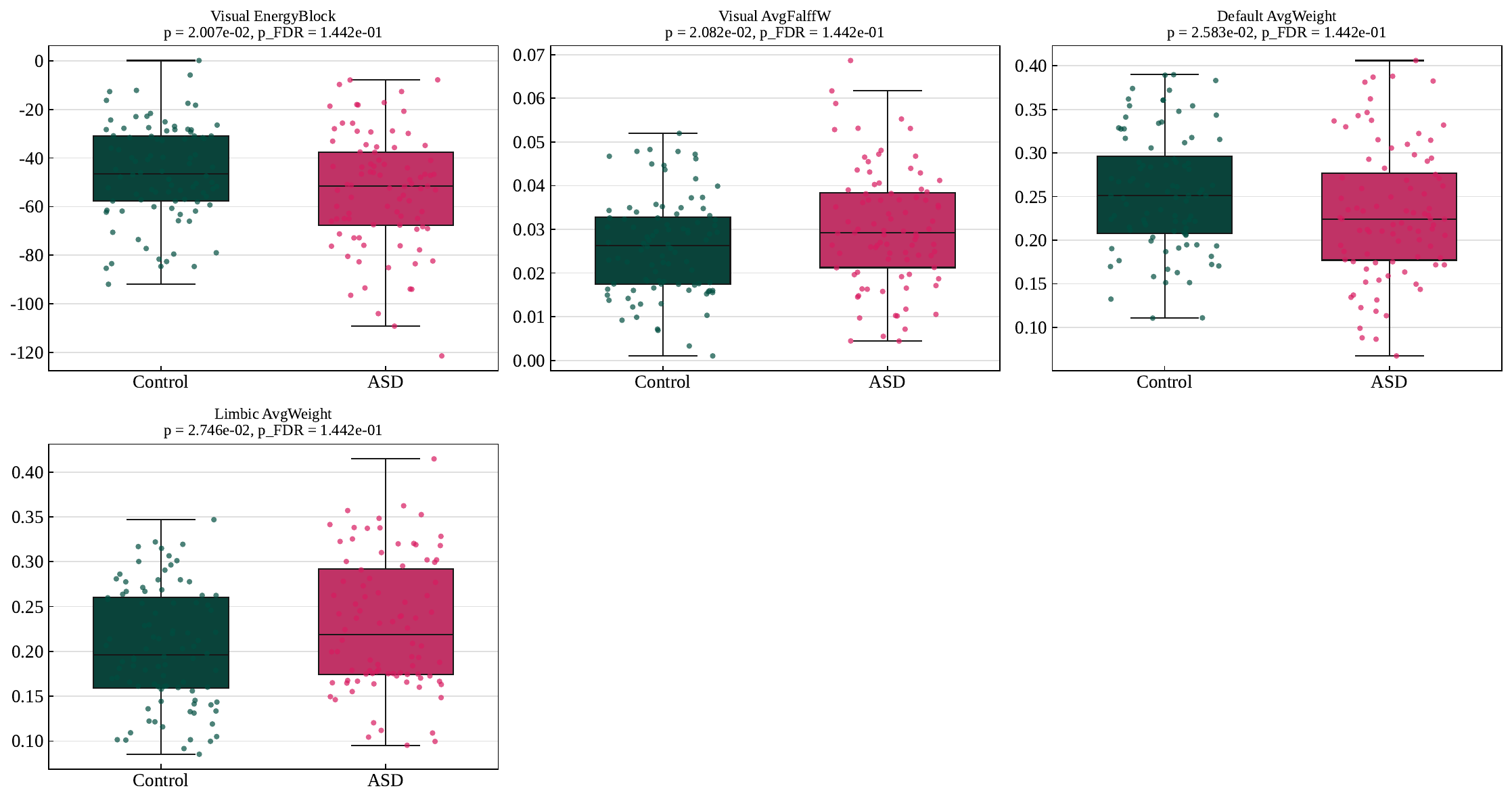}
 \caption{Intra-network metrics showing uncorrected group differences between ASD and TD adults. Boxplots display the distributions for Visual EnergyBlock, Visual AvgFalffW, Default AvgWeight, and Limbic AvgWeight. While these metrics showed nominally significant differences ($p < 0.05$), none survived FDR correction ($p_{\mathrm{FDR}} = 0.144$ for all), and should therefore be interpreted as exploratory findings.}
 \label{fig:subnetworks_FDR-raw}
\end{figure}

\subsection*{Inter-network connectivity and energy}

To extend the intra-network analyses, we next examined how coevolutionary balance and connectivity are organized between large-scale systems. Interactions were quantified for each unordered pair $(u,v)$ of the seven Yeo resting-state networks \cite{Yeo2011}, yielding 21 distinct network pairs. After unified outlier removal based on whole-brain coevolutionary energy (see Methods), 90 autistic and 92 TD participants remained for this analysis.

For every network pair $(u,v)$ we computed three complementary quantities that capture different aspects of between-network coupling: average functional connectivity ($\mathrm{AvgWeight}_{uv}$), fALFF-aligned connectivity ($\mathrm{AvgFalffW}_{uv}$), and inter-network coevolutionary energy ($E_{uv}$), as defined in the Methods.
Group differences in each metric were assessed using both Welch's two-sample $t$-tests and Mann–Whitney $U$ tests applied separately to every network pair and measure. To control for multiple comparisons across the 63 inter-network metrics (21 pairs $\times$ 3 measures), we applied Benjamini–Hochberg FDR correction.
No inter-network metrics survived FDR correction at the conventional $\alpha = 0.05$ threshold. However, several network pairs showed notable uncorrected effects that approached significance after correction. The Visual–Frontoparietal interaction showed the strongest effects, with both AvgFalffW ($p = 0.0022$, $p_{\mathrm{FDR}} = 0.068$) and EnergyBlock ($p = 0.0021$, $p_{\mathrm{FDR}} = 0.068$) trending toward significance. ASD participants showed more positive fALFF-aligned connectivity and more negative inter-network energy for Visual–Frontoparietal interactions, indicating stronger energetically favorable coupling between these systems.
Additional uncorrected effects ($p < 0.05$) were observed for Visual–Salience/VentAttn AvgWeight, Visual–SomatoMotor AvgWeight, Dorsal Attention–Limbic interactions (both AvgFalffW and EnergyBlock), and SomatoMotor–Limbic interactions. Table~\ref{tab:internetwork_new} summarizes the inter-network metrics with the smallest $p$-values.

These findings suggest that, while not reaching the strict FDR threshold, there may be alterations in how visual processing networks interact with frontoparietal control and limbic systems in ASD. The Visual–Frontoparietal effects, in particular, warrant follow-up in larger samples.

\begin{table}[ht]
\centering
\caption{Inter-network differences in average connectivity (AvgWeight), fALFF-aligned connectivity (AvgFalffW), 
and coevolutionary energy (EnergyBlock) between canonical resting-state networks. For each network pair, 
group comparisons were performed on ASD versus TD, and Benjamini–Hochberg FDR correction was applied across all 63 inter-network metrics. 
The table lists all metrics with uncorrected $p < 0.05$. No inter-network metric survived FDR correction at $\alpha = 0.05$; the Visual–Frontoparietal effects ($p_{\mathrm{FDR}} = 0.068$) are treated as exploratory.}
\label{tab:internetwork_new}
\begin{tabular}{llccc}
\toprule
\textbf{Network Pair} & \textbf{Metric} & \textbf{$p$} & \textbf{$p_{\mathrm{FDR}}$} \\
\midrule
Visual $\leftrightarrow$ Frontoparietal & EnergyBlock & $0.0021$ & $0.068$ \\
Visual $\leftrightarrow$ Frontoparietal & AvgFalffW   & $0.0022$ & $0.068$ \\
Visual $\leftrightarrow$ Salience/VentAttn & AvgWeight & $0.013$ & $0.251$ \\
DorsalAttn $\leftrightarrow$ Limbic & EnergyBlock & $0.017$ & $0.251$ \\
Visual $\leftrightarrow$ Default & AvgFalffW & $0.021$ & $0.251$ \\
DorsalAttn $\leftrightarrow$ Limbic & AvgFalffW & $0.022$ & $0.251$ \\
Visual $\leftrightarrow$ SomatoMotor & AvgWeight & $0.024$ & $0.251$ \\
SomatoMotor $\leftrightarrow$ Limbic & AvgFalffW & $0.043$ & $0.289$ \\
SomatoMotor $\leftrightarrow$ Limbic & EnergyBlock & $0.044$ & $0.289$ \\
Visual $\leftrightarrow$ Default & EnergyBlock & $0.045$ & $0.289$ \\
\bottomrule
\end{tabular}
\end{table}

\begin{figure}[H]
 \centering
 \includegraphics[width=0.95\linewidth]{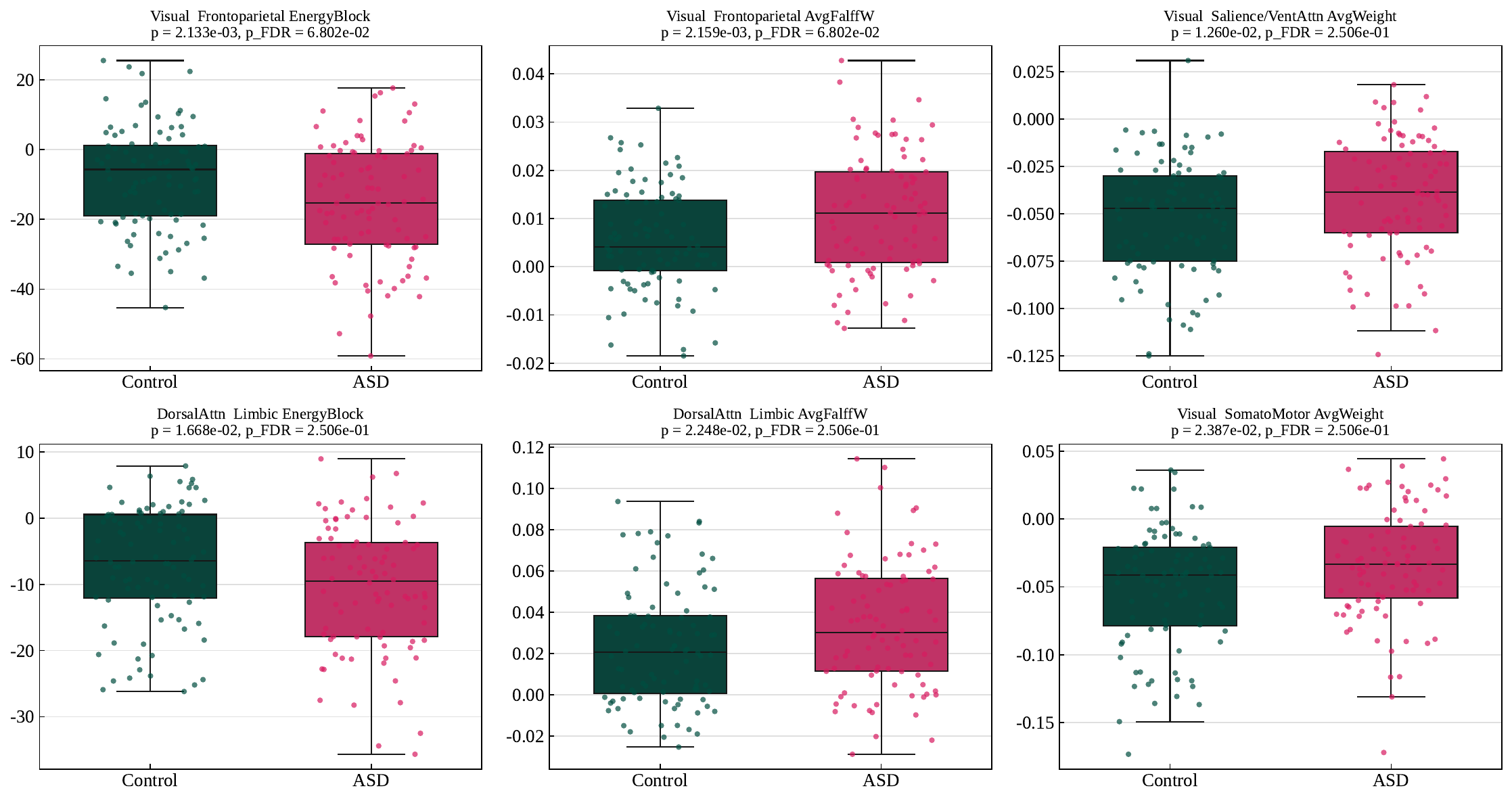}
 \caption{Inter-network metrics showing uncorrected group differences between ASD and TD adults. The top row displays Visual–Frontoparietal EnergyBlock and AvgFalffW, which showed the strongest effects ($p_{\mathrm{FDR}} = 0.068$). Additional panels show Visual–Salience/VentAttn AvgWeight, DorsalAttn–Limbic EnergyBlock and AvgFalffW, and Visual–SomatoMotor AvgWeight. While nominally significant at the uncorrected level, these effects did not survive FDR correction and should be interpreted as exploratory.}
 \label{fig:inter_network}
\end{figure}

\subsection*{Correlation analysis of network metrics with behavioral scores}

We next examined how whole-brain, subnetwork, and inter-network coevolutionary metrics relate to clinical variation in the autistic group. Building on our initial exploratory correlations, we conducted a comprehensive analysis correlating all 90 ComBat-harmonized network features against eight ADI-R and ADOS subscales: ADI-R Social Total A, ADI-R Verbal Total BV, ADI-R RRB Total C, ADI-R Onset Total D, ADOS Total, ADOS Communication, ADOS Social, and ADOS Stereotyped Behaviors. Correlations were computed only in ASD participants with valid clinical scores who were drawn from the same outlier-cleaned cohort used in the energy and connectivity analyses (see Table~\ref{tab:demographics}). This yielded 85 subjects with matched feature and phenotypic data, though sample sizes varied by clinical measure due to missing values (range: $n = 36$--81 per analysis).

\begin{figure}[H]
  \centering
  \includegraphics[width=0.95\linewidth]{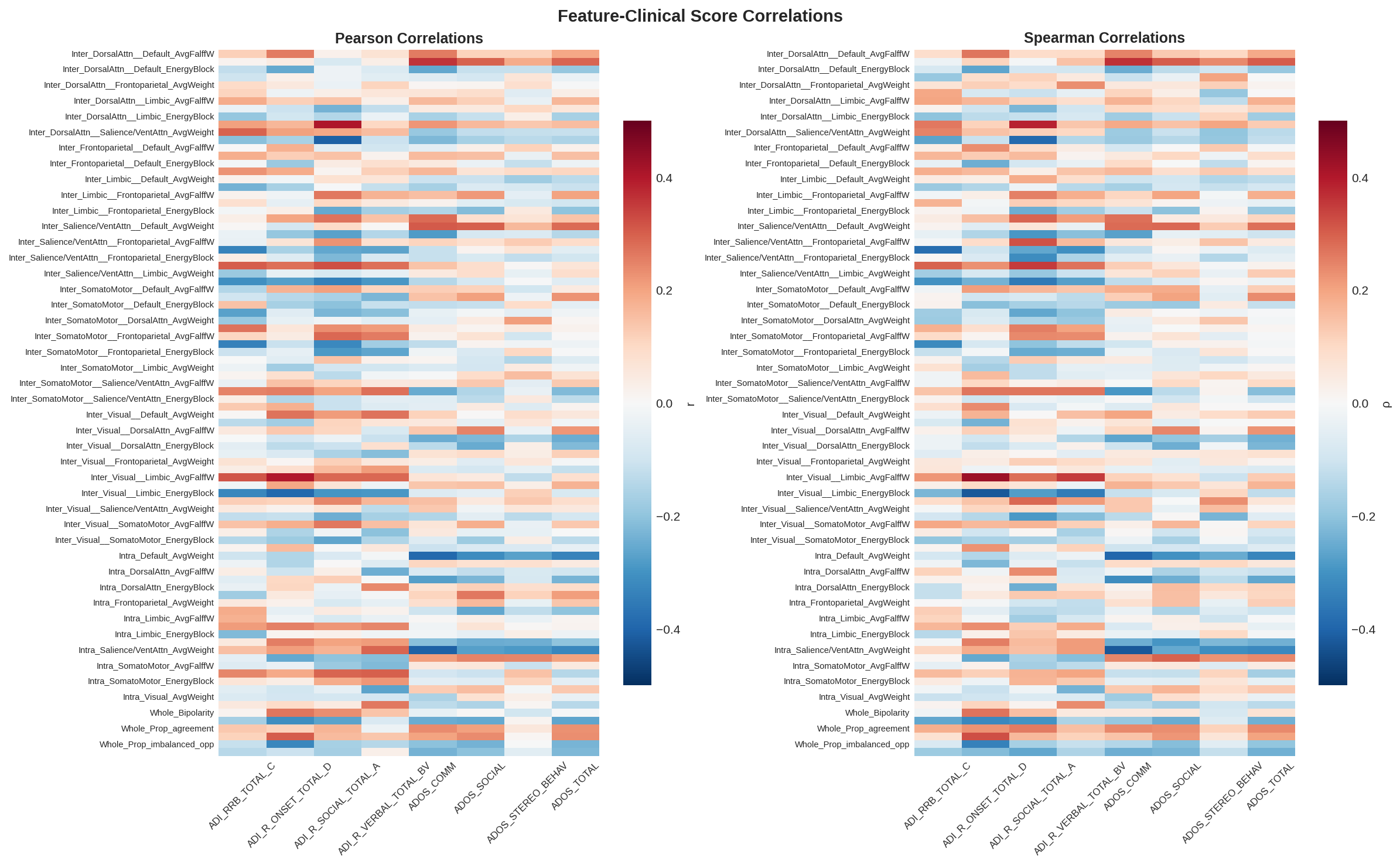}
  \caption{Correlation heatmaps showing the relationship between all 90 network features and 8 clinical scores. Left panel: Pearson correlation coefficients. Right panel: Spearman correlation coefficients. Color scale ranges from $-0.5$ (blue) to $+0.5$ (red). The strongest negative correlations cluster around intra-network connectivity of default mode and salience/ventral attention networks with ADOS measures.}
  \label{fig:corr_heatmap}
\end{figure}

\begin{figure}[ht]
  \centering
  \includegraphics[width=0.95\linewidth]{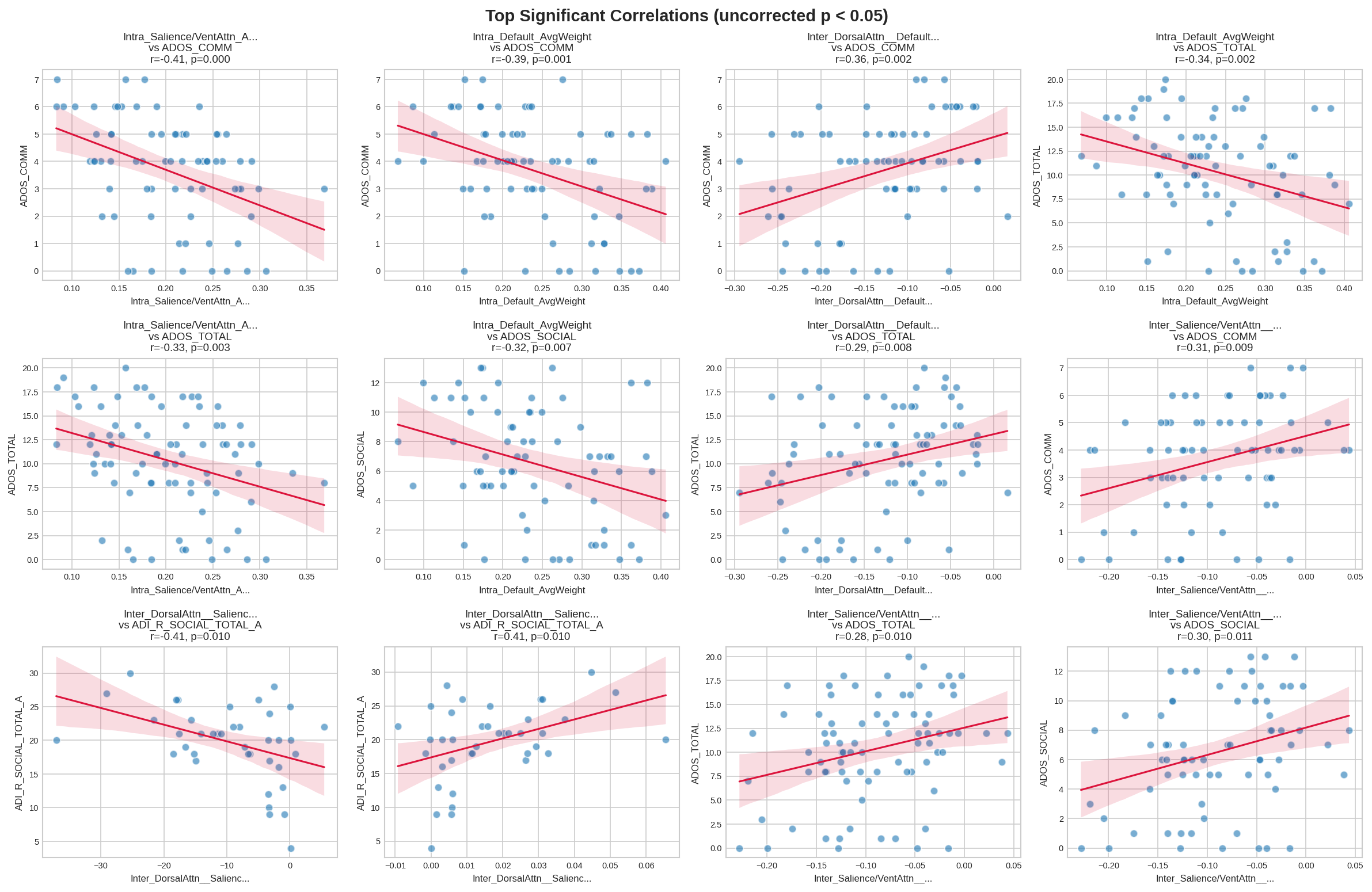}
  \caption{Scatter plots for the top 12 nominally significant correlations between network features and clinical scores (uncorrected $p < 0.05$). Each panel shows the regression line with 95\% confidence interval. The strongest associations involve intra-network connectivity (AvgWeight) of salience/ventral attention and default mode networks with ADOS Communication scores. None of these associations survived FDR correction.}
  \label{fig:corr_scatter}
\end{figure}

For each combination of network metric and behavioral score, we estimated both Pearson's product–moment correlation coefficient ($r$) and Spearman's rank correlation coefficient ($\rho$), together with their exact two-tailed $p$-values. To control for multiple comparisons across the 720 tests (90 features $\times$ 8 clinical scores), we applied Benjamini–Hochberg false discovery rate (FDR) correction separately for Pearson and Spearman $p$-values.
At the uncorrected threshold of $p < 0.05$, 64 feature–clinical score pairs showed nominally significant associations. However, none of these associations survived FDR correction at $q < 0.05$, reflecting the stringent correction required for 720 simultaneous tests with a moderate sample size. The full correlation matrix is provided in Supplementary Table~\ref{tab:supp_corr_full}, and the complete list of nominally significant associations is available in Supplementary Table~\ref{tab:supp_corr_sig}.
Despite the lack of FDR-significant results, a consistent pattern emerged among the strongest uncorrected associations (Table~\ref{tab:behaviour_corr_expanded}). The most robust effects involved intra-network average connectivity (AvgWeight) of the salience/ventral attention and default mode networks correlating negatively with ADOS Communication scores. Specifically, Intra\_Salience/VentAttn\_AvgWeight showed $r = -0.41$, $p = 0.0004$ (Spearman $\rho = -0.42$, $p = 0.0002$; $n = 71$), and Intra\_Default\_AvgWeight showed $r = -0.39$, $p = 0.0007$ (Spearman $\rho = -0.40$, $p = 0.0006$; $n = 71$). These correlations indicate that stronger intra-network functional connectivity within these systems was associated with lower (less severe) ADOS Communication scores.
Inter-network connectivity metrics also showed consistent patterns. The average connectivity between dorsal attention and default mode networks (Inter\_DorsalAttn\_\_Default\_AvgWeight) was positively correlated with ADOS Communication ($r = 0.36$, $p = 0.002$), ADOS Total ($r = 0.29$, $p = 0.008$), and ADOS Social ($r = 0.29$, $p = 0.013$). Similarly, Inter\_Salience/VentAttn\_\_Default\_AvgWeight showed positive associations with all three ADOS measures. These positive correlations suggest that greater inter-network coupling between attention and default mode systems was associated with higher symptom severity.
Whole-brain coevolutionary energy (Whole\_Energy\_global) showed negative correlations with ADOS Total ($r = -0.26$, $p = 0.017$) and ADOS Social ($r = -0.26$, $p = 0.030$), indicating that more negative (energetically favorable) global configurations were modestly associated with lower symptom severity. Several inter-network energy block measures also showed nominally significant associations with ADI-R subscales, including Inter\_DorsalAttn\_\_Salience/VentAttn\_EnergyBlock with ADI-R Social Total A ($r = -0.41$, $p = 0.010$; $n = 39$).
Figure~\ref{fig:corr_heatmap} displays the correlation heatmap for all 90 features against the eight clinical scores, with Figure~\ref{fig:corr_scatter} showing scatter plots for the top associations. The supplementary p-value heatmaps (Supplementary Figure~\ref{fig:supp_pval_heatmap}) illustrate the distribution of uncorrected and FDR-corrected significance across feature–clinical score combinations.

It is important to emphasize that these correlation analyses, while more comprehensive than our initial exploration, remain hypothesis-generating given that no associations survived FDR correction. The consistent directionality of effects across related metrics and clinical measures provides some confidence in the underlying signal, but replication in larger, independent cohorts is essential before drawing firm conclusions about brain–behavior relationships.

\begin{table}[ht]
\centering
\small
\caption{Top associations between network metrics and clinical scores from the comprehensive correlation analysis. The table reports the 15 strongest associations (by uncorrected Pearson $p$-value) from the full analysis of 90 features $\times$ 8 clinical scores (720 tests). For each pair, uncorrected and FDR-corrected $p$-values are shown for both Pearson and Spearman correlations. None survived FDR correction at $q < 0.05$. Full results are provided in Supplementary Tables~\ref{tab:supp_corr_full} and \ref{tab:supp_corr_sig}.}
\label{tab:behaviour_corr_expanded}
\scriptsize
\begin{tabular}{p{4.2cm}lcccccc}
\toprule
\textbf{Network Feature} & \textbf{Clinical Score} & \textbf{$n$} & \textbf{$r$} & \textbf{$p$} & \textbf{$p_{\mathrm{FDR}}$} & \textbf{$\rho$} & \textbf{$p_{\mathrm{FDR}}^{\rho}$} \\
\midrule
Intra\_Sal/VA\_AvgWeight & ADOS\_COMM & 71 & $-$0.41 & 0.0004 & 0.27 & $-$0.42 & 0.17 \\
Intra\_Default\_AvgWeight & ADOS\_COMM & 71 & $-$0.39 & 0.0007 & 0.27 & $-$0.40 & 0.23 \\
Inter\_DA\_\_Def\_AvgWeight & ADOS\_COMM & 71 & 0.36 & 0.002 & 0.39 & 0.36 & 0.43 \\
Intra\_Default\_AvgWeight & ADOS\_TOTAL & 81 & $-$0.34 & 0.002 & 0.39 & $-$0.33 & 0.46 \\
Intra\_Sal/VA\_AvgWeight & ADOS\_TOTAL & 81 & $-$0.33 & 0.003 & 0.41 & $-$0.32 & 0.46 \\
Intra\_Default\_AvgWeight & ADOS\_SOCIAL & 71 & $-$0.32 & 0.007 & 0.64 & $-$0.30 & 0.58 \\
Inter\_DA\_\_Def\_AvgWeight & ADOS\_TOTAL & 81 & 0.29 & 0.008 & 0.64 & 0.30 & 0.58 \\
Inter\_Sal/VA\_\_Def\_AvgWeight & ADOS\_COMM & 71 & 0.31 & 0.009 & 0.64 & 0.28 & 0.58 \\
Inter\_DA\_\_Sal/VA\_EnBl & ADI\_R\_SOC\_A & 39 & $-$0.41 & 0.010 & 0.64 & $-$0.39 & 0.58 \\
Inter\_DA\_\_Sal/VA\_AvgFW & ADI\_R\_SOC\_A & 39 & 0.41 & 0.010 & 0.64 & 0.38 & 0.58 \\
Inter\_Sal/VA\_\_Def\_AvgWeight & ADOS\_TOTAL & 81 & 0.28 & 0.010 & 0.64 & 0.28 & 0.58 \\
Inter\_Sal/VA\_\_Def\_AvgWeight & ADOS\_SOCIAL & 71 & 0.30 & 0.011 & 0.64 & 0.29 & 0.58 \\
Inter\_DA\_\_Def\_AvgWeight & ADOS\_SOCIAL & 71 & 0.29 & 0.013 & 0.64 & 0.30 & 0.58 \\
Inter\_Vis\_\_Lim\_AvgFalffW & ADI\_R\_ONS\_D & 36 & 0.40 & 0.015 & 0.64 & 0.43 & 0.58 \\
Whole\_Energy\_global & ADOS\_TOTAL & 81 & $-$0.26 & 0.017 & 0.64 & $-$0.24 & 0.63 \\
\bottomrule
\end{tabular}
\begin{flushleft}
\footnotesize
Abbreviations: Sal/VA = Salience/VentAttn, DA = DorsalAttn, Def = Default, Vis = Visual, Lim = Limbic, EnBl = EnergyBlock, AvgFW = AvgFalffW, SOC\_A = SOCIAL\_TOTAL\_A, ONS\_D = ONSET\_TOTAL\_D. $p_{\mathrm{FDR}}$ and $p_{\mathrm{FDR}}^{\rho}$ denote Benjamini–Hochberg corrected $p$-values for Pearson and Spearman tests, respectively.
\end{flushleft}
\end{table}

\subsection*{Machine learning classification of ASD versus TD}

To test whether the coevolutionary and connectivity features contain discriminative information about diagnostic status, we trained supervised classifiers to distinguish ASD from TD on the ABIDE I adult male subsample with complete feature data (Table~\ref{tab:demographics}). The analysis was explicitly designed to avoid information leakage and to provide an unbiased estimate of held-out performance.

\paragraph*{Leakage-safe pipeline and evaluation.}
All machine-learning analyses were implemented in Python using \texttt{scikit-learn} and \texttt{XGBoost}. First, the full dataset (participants $\times$ features) was randomly split into an 80\% training set and a 20\% held-out test set using a stratified split, preserving the ASD/TD ratio in both partitions. The test set was \emph{never} used for feature selection, hyperparameter tuning, model comparison, or preprocessing decisions.

Within the training set, we used a nested pipeline in which all preprocessing steps were estimated strictly inside the cross-validation folds. Each model was wrapped in a \texttt{Pipeline} that first applied feature-wise standardization (\texttt{StandardScaler}) and then fitted the classifier. For each split of a stratified five-fold cross-validation, the scaler was fitted \emph{only} on that fold's training portion and then applied to the corresponding validation portion. This guarantees that no information from the validation data leaks into the training transformation. Hyperparameters were tuned with \texttt{GridSearchCV} using the same five-fold stratified scheme and accuracy as the primary scoring metric. The best hyperparameters for each model were selected based solely on mean cross-validated accuracy \emph{within the training set}. After tuning, the best configuration of each classifier was refitted on the entire training set and then evaluated exactly once on the untouched test set, yielding a single unbiased estimate of held-out performance.

We considered five classifiers: support vector machine (SVM, radial basis function kernel), XGBoost gradient boosting, logistic regression, Gaussian Naive Bayes, and $k$-nearest neighbors (KNN). Hyperparameter grids spanned regularization strengths, kernel parameters, tree depths, and neighbor counts in ranges commonly used for small-to-moderate neuroimaging datasets (see Methods for details). Table~\ref{tab:ml_summary_full} summarizes the cross-validated accuracy on the training set, final test accuracy, balanced accuracy, and F1-score for all models.

\begin{table}[ht]
\centering
\caption{Performance of supervised classifiers for ASD versus TD classification using the selected coevolutionary and connectivity features after ComBat harmonization. CV score was obtained from stratified 5-fold cross-validation on the training set (80\% of participants). Test metrics were computed on the held-out 20\% test set ($n = 36$), which was not used at any stage of feature selection or hyperparameter tuning. 95\% confidence intervals (CI) were computed via bootstrap resampling.}
\label{tab:ml_summary_full}
\setlength{\tabcolsep}{4pt}
\small
\begin{tabular}{lcccccc}
\toprule
Model & CV Score & Test Acc & Test Bal Acc & Test F1 & Acc 95\% CI \\
\midrule
Gaussian Naive Bayes  & 0.665 & \textbf{0.778} & \textbf{0.778} & \textbf{0.771} & [0.639, 0.917] \\
SVM (RBF)             & 0.672 & 0.694 & 0.694 & 0.692 & [0.528, 0.833] \\
KNN                   & 0.653 & 0.694 & 0.694 & 0.674 & [0.556, 0.833] \\
XGBoost               & 0.666 & 0.611 & 0.611 & 0.600 & [0.444, 0.778] \\
Logistic Regression   & 0.665 & 0.583 & 0.583 & 0.567 & [0.417, 0.750] \\
\bottomrule
\end{tabular}
\end{table}

Gaussian Naive Bayes achieved the highest held-out test accuracy (77.8\%), balanced accuracy (77.8\%), and F1-score (0.771), followed by SVM and KNN (both 69.4\% accuracy). To better characterize the best-performing model, we computed additional metrics on the held-out test set using its predicted class labels and class probabilities (Table~\ref{tab:ml_best_model_metrics}). The ROC AUC was 0.79, indicating robust above-chance discrimination.

\begin{table}[ht]
\centering
\caption{Detailed performance metrics for the best classifier (Gaussian Naive Bayes) on the held-out test set ($n = 36$). Accuracy and balanced accuracy are reported across both classes; precision, recall, and F1-score refer to the ASD class as the positive class. ROC AUC was computed from the predicted ASD class probabilities.}
\label{tab:ml_best_model_metrics}
\setlength{\tabcolsep}{6pt}
\small
\begin{tabular}{lc}
\toprule
Metric & Value \\
\midrule
Accuracy & 0.778 \\
Balanced accuracy & 0.778 \\
Precision (ASD class) & 0.611 \\
Recall (ASD class) & 0.611 \\
F1-score (macro) & 0.771 \\
ROC AUC & 0.79 \\
\bottomrule
\end{tabular}
\end{table}

\begin{figure}[ht]
  \centering
  \includegraphics[width=0.75\linewidth]{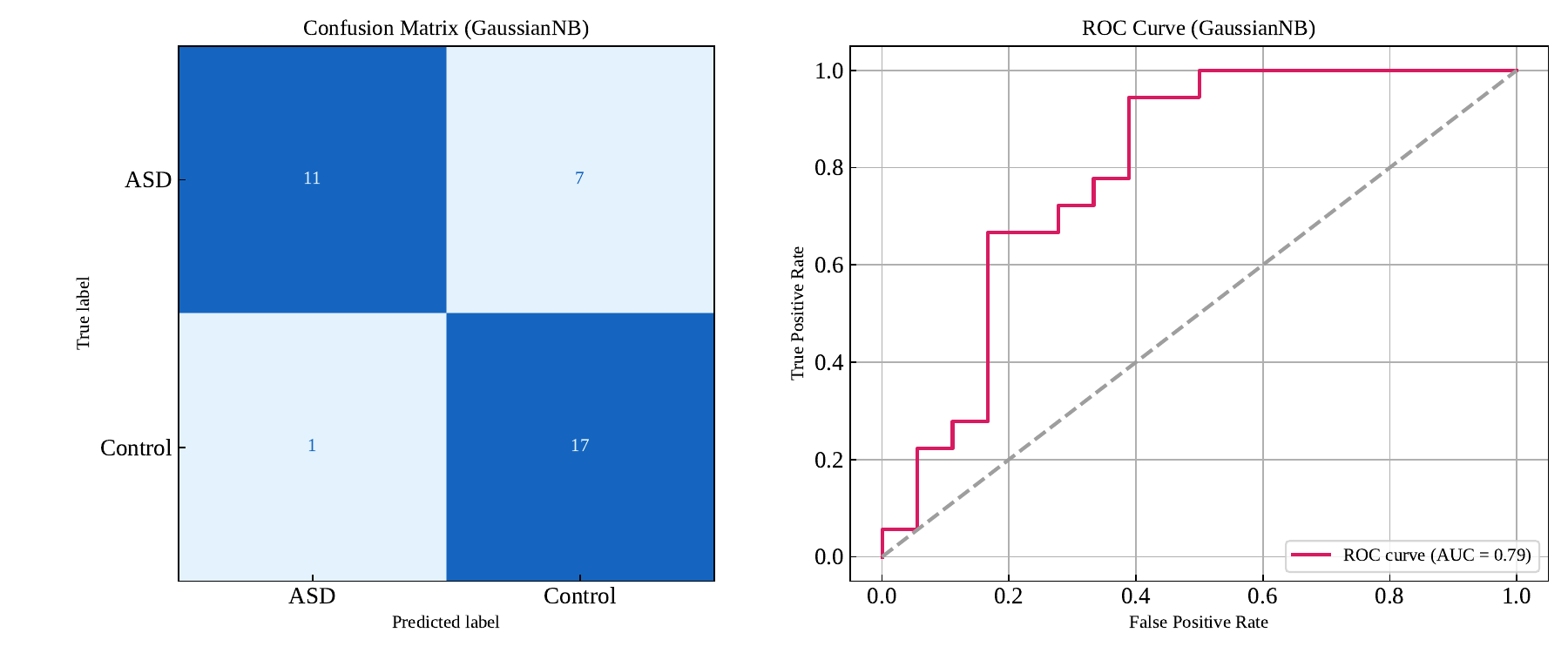}
  \caption{Confusion matrix and receiver operating characteristic (ROC) curve for the best-performing classifier on the held-out test set. The panel on the left shows the confusion matrix for the Gaussian Naive Bayes model trained on the selected coevolutionary and inter-network features, summarizing true and predicted diagnostic labels (ASD versus TD). The panel on the right depicts the corresponding ROC curve based on the predicted probability of the ASD class; the diagonal grey line indicates chance performance. The area under the ROC curve (AUC) was 0.79, indicating robust above-chance discrimination.}
  \label{fig:ml_confusion_roc}
\end{figure}

Figure~\ref{fig:ml_confusion_roc} shows the confusion matrix and ROC curve for this Gaussian Naive Bayes classifier on the test set. The confusion matrix indicates that 11 of 18 ASD participants and 17 of 18 TD participants were correctly classified. The ROC curve lies clearly above the diagonal chance line, yielding an AUC of 0.79, which is comparable to or exceeds recent rs-fMRI classification studies that used carefully leakage-controlled pipelines on ABIDE data \cite{Smith:2013fc}. 

\begin{figure}[H]
  \centering
  \includegraphics[width=0.6\linewidth]{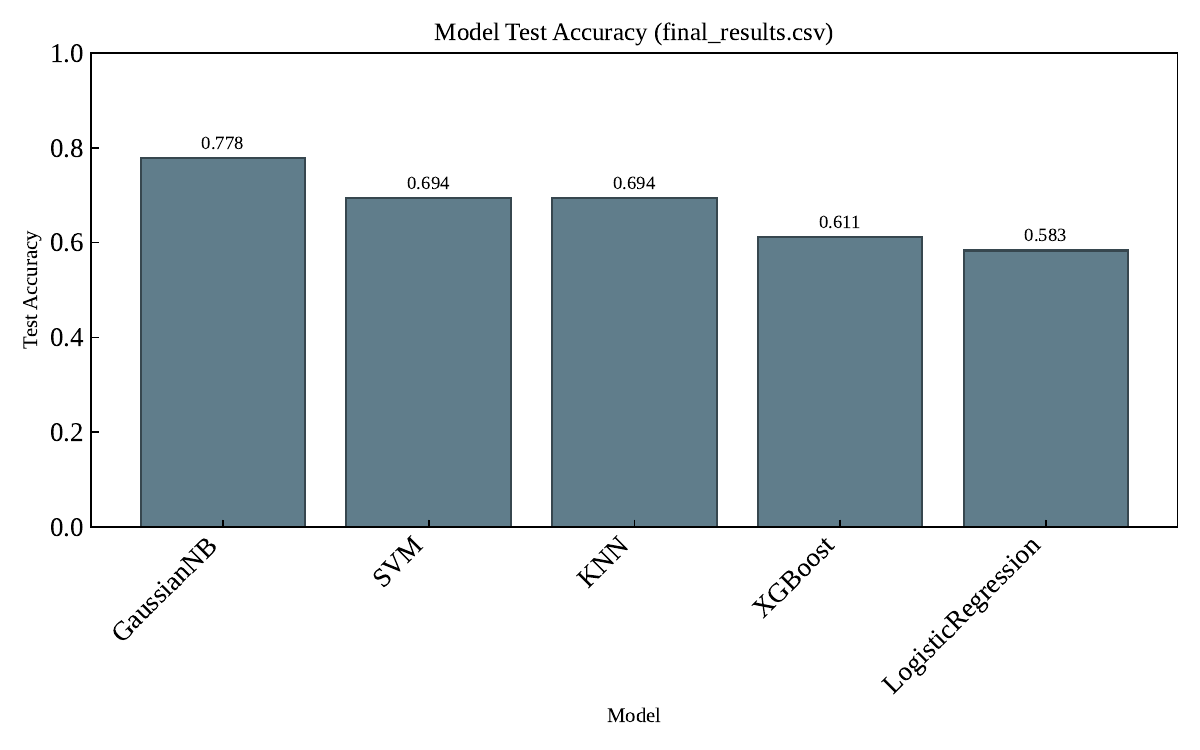}
  \caption{Comparison of test accuracy across all five classifiers. Gaussian Naive Bayes achieved the highest test accuracy (77.8\%), followed by SVM and KNN (both 69.4\%). Error bars represent the range of accuracy values.}
  \label{fig:ml_barplot}
\end{figure}

Figure~\ref{fig:ml_barplot} provides a visual comparison of test accuracy across all five classifiers, highlighting the superior performance of Gaussian Naive Bayes.

\paragraph*{Feature selection and importance-ranked feature set.}
To reduce dimensionality and focus the classifiers on the most informative descriptors, we performed a single feature selection step \emph{before} model-specific training and tuning. This feature selection step used a Random Forest classifier, applied only to the training data, to rank features by mean decrease in impurity. Starting from the full set of 90 graph-based features (whole-brain coevolutionary metrics, intra-network energies, and inter-network connectivity and energy terms), we fitted a random forest classifier on the training set only and extracted the importance for each feature. Features were then ranked by importance, and we retained the top 25\% (23 features) as the final input feature set for all subsequent classifiers.

The full importance ranking is reported in Supplementary Table~\ref{tab:supp_ml_results}. Table~\ref{tab:ml_selected_features} lists the top 25 features by random-forest importance. These features primarily capture inter-network average connectivity involving the Visual, SomatoMotor, Limbic, and Default Mode networks, as well as whole-brain motif proportions, consistent with the results of the univariate analyses.

\begin{table}[H]
\centering
\caption{Top-ranked features contributing to ASD--TD classification. 
The table lists the top 25 graph-based descriptors ranked by random forest feature importance, 
together with their normalized importance scores.}
\label{tab:ml_selected_features}
\setlength{\tabcolsep}{6pt}
\small
\begin{tabular}{clc}
\toprule
Rank & Feature & Importance \\
\midrule
1 & Inter\_Visual\_\_Salience/VentAttn\_AvgWeight & 0.0246 \\
2 & Inter\_Visual\_\_SomatoMotor\_AvgWeight & 0.0241 \\
3 & Whole\_Prop\_imbalanced\_same & 0.0230 \\
4 & Whole\_Prop\_agreement & 0.0215 \\
5 & Intra\_DorsalAttn\_AvgWeight & 0.0201 \\
6 & Inter\_SomatoMotor\_\_Limbic\_AvgWeight & 0.0191 \\
7 & Inter\_Limbic\_\_Default\_EnergyBlock & 0.0191 \\
8 & Inter\_DorsalAttn\_\_Default\_AvgWeight & 0.0183 \\
9 & Intra\_Limbic\_EnergyBlock & 0.0175 \\
10 & Intra\_Visual\_EnergyBlock & 0.0168 \\
11 & Inter\_DorsalAttn\_\_Limbic\_AvgFalffW & 0.0155 \\
12 & Inter\_Limbic\_\_Default\_AvgWeight & 0.0154 \\
13 & Inter\_Visual\_\_Frontoparietal\_AvgFalffW & 0.0153 \\
14 & Intra\_Limbic\_AvgFalffW & 0.0153 \\
15 & Inter\_DorsalAttn\_\_Limbic\_EnergyBlock & 0.0152 \\
16 & Inter\_Frontoparietal\_\_Default\_AvgFalffW & 0.0149 \\
17 & Inter\_SomatoMotor\_\_Default\_AvgWeight & 0.0147 \\
18 & Inter\_Visual\_\_Frontoparietal\_EnergyBlock & 0.0144 \\
19 & Inter\_Visual\_\_Limbic\_AvgFalffW & 0.0143 \\
20 & Inter\_DorsalAttn\_\_Frontoparietal\_AvgFalffW & 0.0142 \\
21 & Inter\_SomatoMotor\_\_Limbic\_EnergyBlock & 0.0139 \\
22 & Intra\_Visual\_AvgFalffW & 0.0137 \\
23 & Inter\_Limbic\_\_Default\_AvgFalffW & 0.0137 \\
24 & Inter\_Salience/VentAttn\_\_Frontoparietal\_AvgWeight & 0.0135 \\
25 & Inter\_Frontoparietal\_\_Default\_AvgWeight & 0.0135 \\
\bottomrule
\end{tabular}
\end{table}

Overall, the classification results reinforce the main univariate findings: features that describe how coevolutionary energy and signed connectivity are distributed within and between visual, limbic, default mode, and attention networks carry the bulk of the diagnostic signal, while whole-brain motif proportions (agreement and imbalanced-same) also contribute substantially to ASD versus TD discrimination in this cohort.

\subsection*{Sensitivity analysis: results without global signal regression}

Because GSR can introduce artifactual negative correlations \cite{Murphy2009,Fox2009,Chang2009} and the sign of functional links is a key parameter in the coevolutionary framework, we repeated all primary analyses on data preprocessed without GSR (CompCor-based denoising only; see Methods). After identical quality control and Tukey IQR-based outlier removal, the no-GSR sample comprised 80 ASD and 88 TD participants.

\begin{figure}[H]
 \centering
 \includegraphics[width=0.55\linewidth]{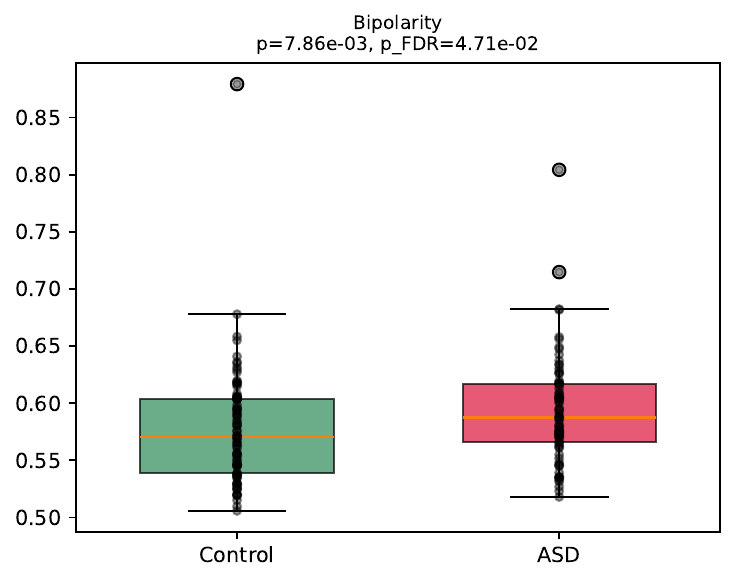}
 \caption{Whole-brain bipolarity in the no-GSR sensitivity analysis. Bipolarity was the only whole-brain metric that survived FDR correction without GSR ($p_{\mathrm{FDR}} = 0.047$), with ASD showing higher bipolarity (stronger two-block organization) than TD. This contrasts with the GSR-based analysis, where bipolarity did not differ between groups.}
 \label{fig:noGSR_whole}
\end{figure}

\paragraph*{Whole-brain metrics.}
In contrast to the GSR-based analysis, global coevolutionary energy did not differ significantly between groups in the no-GSR pipeline (ASD: $-284.4 \pm 151.7$; TD: $-308.4 \pm 164.6$; Mann--Whitney $U$, $p = 0.355$). Similarly, the proportions of agreement, disagreement, imbalanced-same, and imbalanced-opposite links did not reach significance after FDR correction (all $p_{\mathrm{FDR}} > 0.09$; Table~\ref{tab:noGSR_whole}). However, whole-brain bipolarity, which did not differ between groups in the primary GSR analysis ($p = 0.42$), showed a significant group difference without GSR (ASD: $0.593 \pm 0.046$; TD: $0.576 \pm 0.050$; $p = 0.008$, $p_{\mathrm{FDR}} = 0.047$; Figure~\ref{fig:noGSR_whole}), with the ASD group exhibiting higher bipolarity, indicating a stronger two-block organization of the signed network when negative correlations are not induced by GSR.

\begin{table}[ht]
\centering
\caption{Whole-brain coevolutionary metrics in the no-GSR sensitivity analysis. The table reports ASD and TD group means, standard deviations, the statistical test used, raw $p$-values, and FDR-corrected $p$-values. Only bipolarity survived FDR correction ($p_{\mathrm{FDR}} = 0.047$). Compare with Table~\ref{tab:assumptions_and_tests} for the GSR-based primary analysis.}
\label{tab:noGSR_whole}
\small
\resizebox{\columnwidth}{!}{%
\begin{tabular}{lccccc}
\toprule
Metric & ASD Mean (SD) & TD Mean (SD) & Test & $p$ & $p_{\mathrm{FDR}}$ \\
\midrule
Energy global           & $-284.4$ (151.7)  & $-308.4$ (164.6)  & Mann--Whitney $U$ & $0.355$ & $0.355$ \\
Bipolarity              & $0.593$ (0.046)   & $0.576$ (0.050)   & Mann--Whitney $U$ & $0.008$ & $0.047$ \\
Prop agreement          & $0.430$ (0.044)   & $0.443$ (0.039)   & Mann--Whitney $U$ & $0.065$ & $0.095$ \\
Prop imbalanced same    & $0.067$ (0.044)   & $0.055$ (0.039)   & Mann--Whitney $U$ & $0.065$ & $0.095$ \\
Prop disagreement       & $0.072$ (0.047)   & $0.059$ (0.040)   & Mann--Whitney $U$ & $0.079$ & $0.095$ \\
Prop imbalanced opp     & $0.430$ (0.047)   & $0.444$ (0.040)   & Mann--Whitney $U$ & $0.079$ & $0.095$ \\
\bottomrule
\end{tabular}%
}
\end{table}

\begin{figure}[H]
 \centering
 \includegraphics[width=0.55\linewidth]{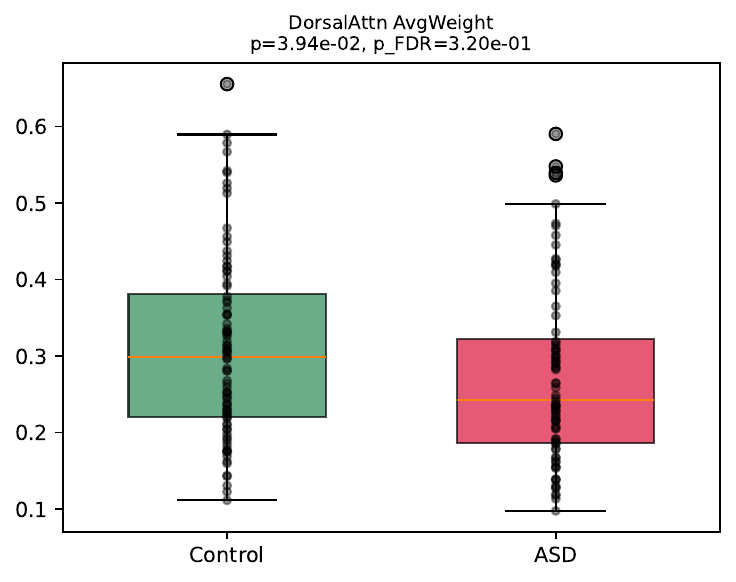}
 \caption{Intra-network metric showing the largest uncorrected group difference in the no-GSR analysis: Dorsal Attention network average connectivity ($p = 0.039$, $p_{\mathrm{FDR}} = 0.32$). This did not survive FDR correction.}
 \label{fig:noGSR_intra}
\end{figure}

\paragraph*{Intra-network metrics.}
At the intra-network level, only Dorsal Attention network average connectivity (AvgWeight) showed an uncorrected group difference (ASD: $0.274 \pm 0.118$; TD: $0.310 \pm 0.121$; $p = 0.039$), with lower connectivity in ASD, but this did not survive FDR correction ($p_{\mathrm{FDR}} = 0.32$; Figure~\ref{fig:noGSR_intra}). No other intra-network metrics reached uncorrected significance.

\begin{figure}[H]
 \centering
 \includegraphics[width=0.95\linewidth]{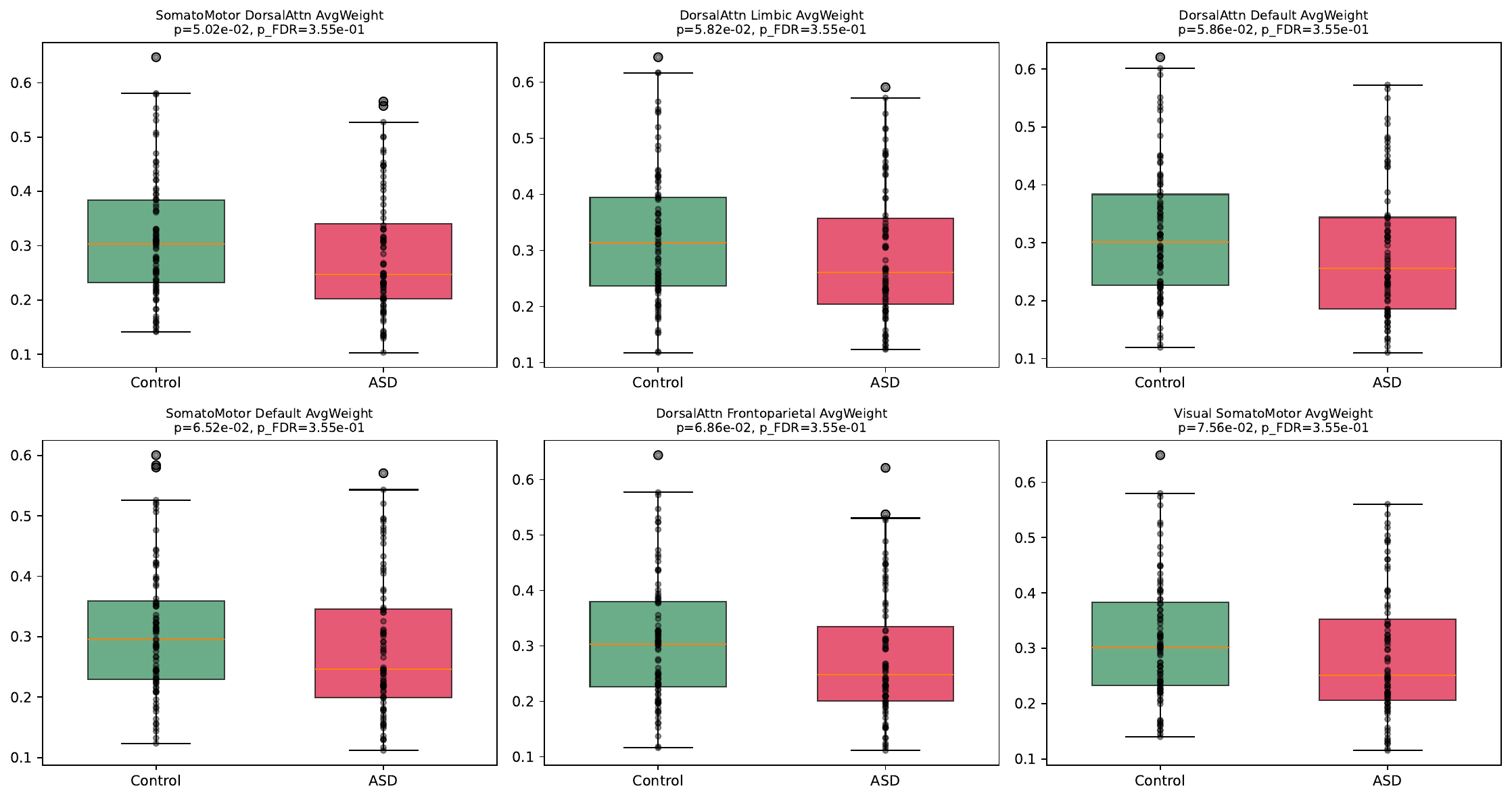}
 \caption{Inter-network average connectivity metrics with the smallest $p$-values in the no-GSR sensitivity analysis. None survived FDR correction (all $p_{\mathrm{FDR}} > 0.35$). Effects are generally weaker and fewer than in the GSR-based analysis.}
 \label{fig:noGSR_inter}
\end{figure}

\paragraph*{Inter-network metrics.}
No inter-network metrics reached significance after FDR correction in the no-GSR analysis (all $p_{\mathrm{FDR}} > 0.35$). Several network pairs involving the dorsal attention and somatomotor networks showed nominally significant trends in average connectivity (SomatoMotor--DorsalAttn: $p = 0.050$; DorsalAttn--Limbic: $p = 0.058$; DorsalAttn--Default: $p = 0.059$; Figure~\ref{fig:noGSR_inter}), consistent with the general pattern of lower connectivity in ASD observed in the GSR analysis.

\begin{figure}[H]
  \centering
  \begin{minipage}{0.48\textwidth}
    \centering
    \includegraphics[width=\linewidth]{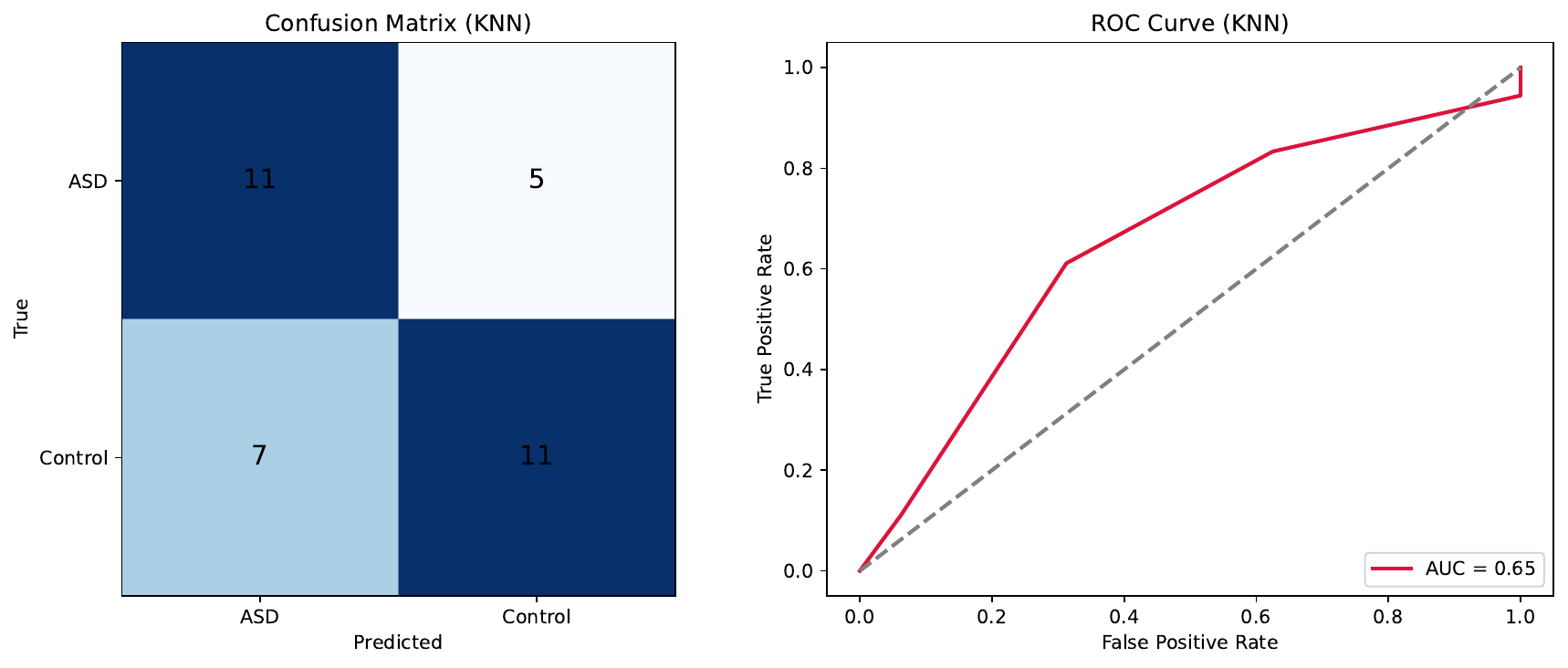}
  \end{minipage}\hfill
  \begin{minipage}{0.48\textwidth}
    \centering
    \includegraphics[width=\linewidth]{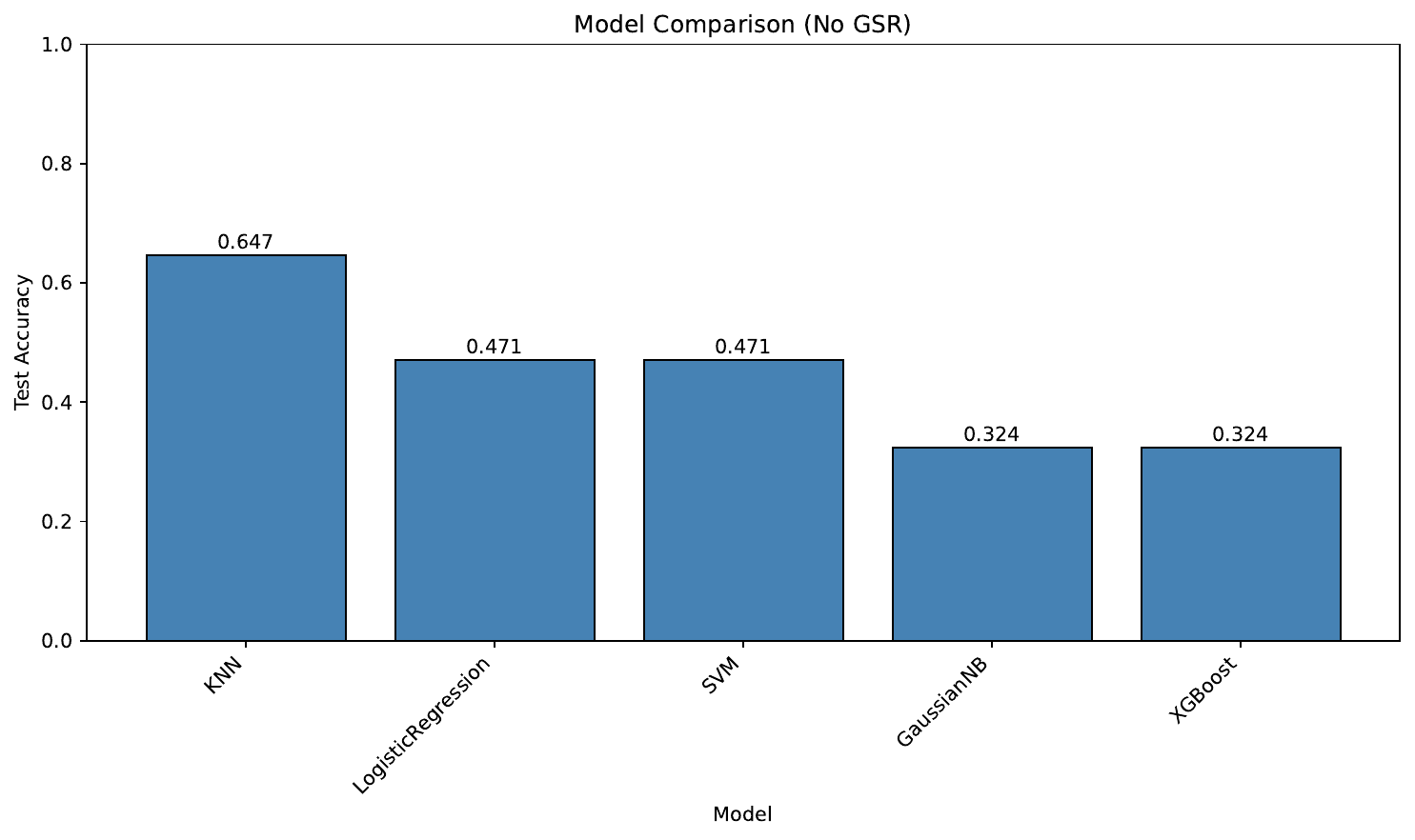}
  \end{minipage}
  \caption{Machine learning results for the no-GSR sensitivity analysis. Left: confusion matrix and ROC curve for the best model (KNN, accuracy = 64.7\%, AUC = 0.65). Right: comparison of test accuracy across all five classifiers. Performance is substantially lower than in the GSR-based primary analysis (Figure~\ref{fig:ml_confusion_roc}).}
  \label{fig:noGSR_ml}
\end{figure}

\paragraph*{Machine learning classification.}
Classification performance was substantially reduced without GSR. The best-performing model was KNN (with $k = 5$), achieving 64.7\% test accuracy (balanced accuracy: 64.9\%; AUC = 0.65) on a held-out test set of $n = 34$ (Table~\ref{tab:noGSR_ml}; Figure~\ref{fig:noGSR_ml}). This represents a notable decrease from the 77.8\% accuracy (AUC = 0.79) obtained with GSR, suggesting that the additional negative correlations introduced by GSR, while potentially artifactual, amplified the discriminative signal captured by the coevolutionary features.

\begin{table}[ht]
\centering
\caption{Classification performance in the no-GSR sensitivity analysis. KNN achieved the highest test accuracy (64.7\%). Compare with Table~\ref{tab:ml_summary_full} for the GSR-based primary analysis.}
\label{tab:noGSR_ml}
\small
\begin{tabular}{lcccc}
\toprule
Model & CV Score & Test Acc & Test Bal Acc & Test F1 \\
\midrule
KNN                   & 0.627 & \textbf{0.647} & \textbf{0.649} & \textbf{0.647} \\
Logistic Regression   & 0.625 & 0.471 & 0.479 & 0.463 \\
SVM (RBF)             & 0.626 & 0.471 & 0.479 & 0.463 \\
Gaussian Naive Bayes  & 0.589 & 0.324 & 0.333 & 0.309 \\
XGBoost               & 0.611 & 0.324 & 0.323 & 0.323 \\
\bottomrule
\end{tabular}
\end{table}

\paragraph*{Summary of GSR sensitivity analysis.}
Table~\ref{tab:GSR_comparison} provides a side-by-side comparison of key results across the two preprocessing pipelines. The divergent patterns across pipelines indicate that some of the effects observed in the primary analysis are sensitive to GSR. In particular, the whole-brain energy and motif proportion differences that were FDR-significant with GSR were attenuated without it, likely because GSR-induced negative correlations amplify the signed structure that the coevolutionary Hamiltonian captures. Conversely, the emergence of bipolarity as the sole FDR-significant metric without GSR suggests that the two-block organization of the network may be a more robust signature of ASD when artifactual anticorrelations are minimized. These complementary findings are discussed further below.

\begin{table}[ht]
\centering
\caption{Comparison of key results between the primary (with GSR) and sensitivity (without GSR) analyses. The table highlights how preprocessing choice affects the main findings.}
\label{tab:GSR_comparison}
\small
\begin{tabular}{lcc}
\toprule
\textbf{Finding} & \textbf{With GSR} & \textbf{Without GSR} \\
\midrule
Sample size (ASD / TD) & 90 / 92 & 80 / 88 \\
\midrule
\multicolumn{3}{l}{\textit{Whole-brain metrics (FDR-significant)}} \\
\quad Energy global & $p_{\mathrm{FDR}} = 0.002$ & $p = 0.355$ (n.s.) \\
\quad Prop agreement & $p_{\mathrm{FDR}} = 0.002$ & $p = 0.065$ (n.s.) \\
\quad Prop imbalanced same & $p_{\mathrm{FDR}} = 0.002$ & $p = 0.065$ (n.s.) \\
\quad Bipolarity & $p = 0.42$ (n.s.) & $p_{\mathrm{FDR}} = 0.047$ \\
\midrule
\multicolumn{3}{l}{\textit{Intra-/Inter-network metrics}} \\
\quad FDR-significant effects & None & None \\
\midrule
\multicolumn{3}{l}{\textit{Classification (best model)}} \\
\quad Model & Gaussian NB & KNN \\
\quad Accuracy & 77.8\% & 64.7\% \\
\quad AUC & 0.79 & 0.65 \\
\bottomrule
\end{tabular}
\end{table}

\section*{Discussion}

In this study we used a coevolutionary balance framework to characterize how resting-state functional networks are organized in adult males with autism spectrum disorder. By combining binarized regional activity (fractional amplitude of low-frequency fluctuations, fALFF) with signed functional connectivity (FC) in a single Hamiltonian, we derived multi-scale energy-based descriptors that jointly capture nodal states and link signs \cite{Heider:1946ac,cartwright1956,Kargaran:2021he,GhanbarzadehNoudehi2022}. We then examined these descriptors at whole-brain, intra-network, and inter-network levels, evaluated them against topology-preserving null models, related them to clinical measures, and tested their utility in supervised classification. Critically, we applied ComBat harmonization to mitigate multi-site batch effects \cite{Fortin2017,Johnson2007} and repeated all analyses with and without GSR to assess the sensitivity of our signed connectivity measures to this preprocessing choice. Overall, the findings suggest that ASD is associated with altered signed network organization at the whole-brain level, though the specific metric capturing this alteration, coevolutionary energy and motif proportions (with GSR) versus bipolarity (without GSR), depends on the denoising strategy used.

At the whole-brain level, autistic participants showed significantly more negative coevolutionary energy than typically developing controls (ASD: $-264.9 \pm 110.3$; TD: $-210.9 \pm 83.5$; $p_{\mathrm{FDR}} = 1.95 \times 10^{-3}$), indicating a shift toward energetically more favorable configurations in which highly active regions are positively coupled and less active regions are negatively coupled. This effect was robust to FDR correction following ComBat harmonization.

Motif-based descriptors derived from the same Hamiltonian showed robust ASD–TD differences after FDR correction in the GSR-based analysis. Autistic networks exhibited a higher proportion of ``agreement'' links and a lower proportion of imbalanced-same links ($p_{\mathrm{FDR}} < 0.002$ for both). Additionally, disagreement and imbalanced-opposite proportions showed significant group differences ($p_{\mathrm{FDR}} = 0.01$). Bipolarity did not differ reliably between groups in the GSR analysis, though it emerged as the sole FDR-significant metric in the no-GSR sensitivity analysis ($p_{\mathrm{FDR}} = 0.047$; see below). Taken together, these results indicate that ASD is characterized less by a loss of global two-block balance and more by a systematic redistribution of local signed motifs. Specifically, autistic networks show a higher proportion of energetically favorable (agreement) link configurations and a lower proportion of unfavorable (imbalanced-same) configurations relative to controls. Within the coevolutionary framework, this redistribution corresponds to the ASD group occupying a deeper energy minimum in the Hamiltonian landscape, that is, a network state in which nodal activity and signed connectivity are more tightly aligned, rather than reflecting a change in the overall modular (bipolar) organization of the network.

When we projected the networks onto canonical Yeo systems \cite{Yeo2011}, intra-network coevolutionary energy did not show statistically robust group differences after FDR correction (all $p_{\mathrm{FDR}} > 0.14$). The Visual network exhibited the largest uncorrected effects, with more negative intra-network energy and higher fALFF-aligned connectivity in ASD than TD ($p \approx 0.02$), and the Default Mode and Limbic networks showed trends in average connectivity. These intra-network findings therefore remain exploratory, suggesting that coevolutionary alterations in ASD are expressed most clearly at the whole-brain motif level rather than within single canonical networks.

The inter-network analyses similarly did not reveal FDR-significant effects. The strongest trends involved the Visual–Frontoparietal interaction ($p_{\mathrm{FDR}} = 0.068$), where ASD showed more positive fALFF-aligned connectivity and more negative inter-network energy. Additional uncorrected effects involved Dorsal Attention–Limbic and SomatoMotor–Limbic interactions. While these patterns did not reach the strict FDR threshold, they provide hypothesis-generating evidence for altered coupling between sensory processing and control/limbic systems in autism that warrants follow-up in larger, independent samples.

Our comprehensive correlational analysis of 90 network features against eight ADI-R and ADOS subscales (720 tests) identified 64 nominally significant associations at uncorrected $p < 0.05$, though none survived Benjamini–Hochberg FDR correction. The strongest uncorrected effects involved intra-network average connectivity of the salience/ventral attention network ($r = -0.41$, $p = 0.0004$) and default mode network ($r = -0.39$, $p = 0.0007$) correlating negatively with ADOS Communication scores. These findings suggest that stronger within-network functional coherence in these systems may be associated with milder communication-related autistic features. Conversely, inter-network connectivity between dorsal attention and default mode networks showed positive correlations with ADOS measures ($r = 0.29$--$0.36$, $p = 0.002$--$0.013$), indicating that greater between-network coupling was associated with higher symptom severity. Whole-brain coevolutionary energy showed modest negative correlations with ADOS Total and Social scores ($r \approx -0.26$, $p < 0.03$). While the lack of FDR-significant results reflects the stringent correction required for 720 simultaneous tests with moderate sample size ($n = 36$--$81$), the consistent directionality of effects across related metrics provides hypothesis-generating evidence that warrants replication in larger cohorts.

The supervised classification analysis demonstrated that coevolutionary features contain robust diagnostic information. Using a leakage-safe pipeline with stratified 80/20 train–test split, Gaussian Naive Bayes achieved 77.8\% test accuracy with an ROC AUC of 0.79. This performance compares favorably with recent methodologically careful rs-fMRI classification studies on ABIDE data that applied similarly strict leakage controls. Without GSR, classification accuracy dropped to 64.7\% (KNN; AUC = 0.65), suggesting that the discriminative signal is partly dependent on the signed connectivity structure enhanced by GSR. The features driving classification, inter-network connectivity involving Visual, Limbic, Default Mode, and attention networks, together with whole-brain motif proportions, are directly interpretable in terms of coevolutionary balance. Importantly, feature selection was performed strictly within the training set, ensuring unbiased performance estimates.

From a methodological standpoint, the application of ComBat harmonization proved essential. The original ABIDE I dataset comprises data from 14 different sites with varying scanner hardware, acquisition parameters, and participant demographics. Without harmonization, site-related variance can inflate or obscure true biological effects. ComBat uses an empirical Bayes framework to estimate and remove site effects while preserving biological variability of interest \cite{Johnson2007,Fortin2017}. Following harmonization, our whole-brain effects became significant at the FDR level, and classification performance improved, suggesting that batch effect removal enhanced our ability to detect genuine ASD-related alterations.

Our sensitivity analysis without GSR revealed that the choice of denoising strategy has important implications for the coevolutionary balance results. GSR is known to shift the distribution of functional correlations toward zero and to introduce artifactual negative correlations \cite{Murphy2009,Fox2009,Chang2009}, which directly affects the sign structure of the functional connectivity matrix. Since the coevolutionary Hamiltonian explicitly depends on signed edge weights, GSR-induced anticorrelations can amplify the signed structure that the energy measure captures, potentially inflating group differences in energy and motif proportions. Consistent with this, the robust whole-brain energy and motif differences observed with GSR ($p_{\mathrm{FDR}} < 0.01$) were attenuated to non-significance without it. Conversely, bipolarity, which measures the global two-block organization of the signed network independently of the coevolutionary energy, emerged as the sole FDR-significant metric without GSR ($p_{\mathrm{FDR}} = 0.047$), with ASD showing higher bipolarity. This finding suggests that autistic brains may exhibit a more clearly polarized modular structure when the analysis is not confounded by GSR-induced anticorrelations. Classification accuracy also declined substantially without GSR (from 77.8\% to 64.7\%), indicating that the discriminative power of coevolutionary features is partly dependent on the enhanced signed structure introduced by GSR. Together, these results indicate that the core finding of altered signed network organization in ASD is present under both preprocessing regimes, but the specific metric that captures this alteration differs: energy and motif proportions with GSR, and bipolarity without GSR. This dissociation underscores the importance of reporting results under both conditions when working with signed connectivity measures \cite{Murphy2009,Fox2009}.

The coevolutionary balance framework offers a principled way to integrate nodal activity and signed connectivity into a single energetic description. Classical graph-theoretic analyses typically treat node-wise measures and edge-wise measures separately. In contrast, the present Hamiltonian explicitly encodes how nodal states and link signs co-align, thereby capturing higher-order constraints invisible to pairwise FC alone. Within this view, the ASD–TD differences observed here can be interpreted as evidence that autistic brains occupy alternative minima in a shared coevolutionary landscape, characterized by shifts in the distribution of local signed motifs.

\paragraph*{Limitations and future directions.}
The present analyses were restricted to adult males from ABIDE I and therefore do not generalize to females, younger individuals, or autistic individuals with intellectual disability. Despite ComBat harmonization, residual site-related variability cannot be fully excluded, and the inherent heterogeneity of ABIDE in scanner hardware and behavioral protocols remains a limitation. As demonstrated by our sensitivity analysis, GSR substantially affects the sign structure of functional connectivity and consequently the coevolutionary energy measures derived from it; the primary GSR-based results should therefore be interpreted with this caveat in mind, and the no-GSR results provide an important complementary perspective. The binarization of fALFF into high versus low values using a subject-specific median threshold simplifies the underlying continuous distribution, and the use of static functional connectivity collapses potentially important temporal dynamics. The choice of the CC200 parcellation and Pearson correlation–based signed connectivity may also influence the derived energy metrics; future work should examine robustness across parcellations, preprocessing strategies, and dynamic connectivity estimators. The cross-sectional design limits inferences about developmental trajectories or causal mechanisms. Finally, while our machine-learning results are encouraging, the modest test set sizes ($n = 34$--$36$) warrant caution, and validation in larger independent datasets is essential.

Future work should address these limitations through longitudinal and developmental designs to test whether atypical trajectories of coevolutionary energy predict later symptom profiles or treatment response. Time-resolved extensions could relate dynamic changes in the Hamiltonian to moment-to-moment fluctuations in network states. Integrating coevolutionary descriptors with genetic, cognitive, and behavioral data may help carve the heterogeneity of ASD into subtypes defined by distinct patterns of large-scale network balance. These directions highlight that energy-based descriptions are a complementary tool that can provide a physically grounded perspective on how large-scale brain networks are organized in neurodevelopmental conditions such as autism.
 
\section*{Methods}

\subsection*{Participants and Dataset}
We obtained resting-state fMRI data from the Autism Brain Imaging Data Exchange I (ABIDE I) repository \cite{DiMartino:2014abide}. A priori power analysis in G*Power (two-tailed, $\alpha=0.05$, $1-\beta=0.80$, $d=0.50$) showed that there needed to be at least 64 individuals in each group \cite{Faul2009}. We randomly selected 100 males with a clinical diagnosis of ASD and 100 age- and IQ-matched TD men who were all right-handed, between 18 and 30 years of age, and had a full-scale IQ greater than 80. This was done to allow for the removal of outliers and site-specific variability. Data were drawn from 14 contributing sites. Certified clinicians at each site used the Autism Diagnostic Observation Schedule (ADOS) and the Autism Diagnostic Interview–Revised (ADI-R) to confirm ASD diagnosis according to DSM-5 criteria \cite{APA:2022ds}.

\begin{table}[H]
\centering
\caption{Demographic and sampling characteristics of the study cohort.}
\label{tab:demographics}
\setlength{\tabcolsep}{3pt}
\small
\resizebox{\columnwidth}{!}{%
\begin{tabular}{p{0.38\columnwidth}p{0.27\columnwidth}p{0.27\columnwidth}}
\toprule
\textbf{Characteristic} & \textbf{ASD} & \textbf{TD} \\
\midrule
Initially selected from ABIDE I, $n$ & 100 & 100 \\
Analyzed sample after quality control, $n$ & 90 & 92 \\
Sex (male / female), $n$ & 90 / 0 & 92 / 0 \\
Age range (years) & 18--30 & 18--30 \\
Handedness (right / left), $n$ & 90 / 0 & 92 / 0 \\
Full-scale IQ (FSIQ) inclusion & $>$ 80 & $>$ 80 \\
Diagnostic status & DSM-5 ASD, ADOS/ADI-R confirmed & Typically developing (TD), no ASD diagnosis \\
Data source / preprocessing & ABIDE I, CPAC pipeline, CC200 atlas (200 ROIs) & ABIDE I, CPAC pipeline, CC200 atlas (200 ROIs) \\
Number of contributing sites & 14 & 14 \\
\bottomrule
\end{tabular}%
}
\begin{flushleft}
\footnotesize
All participants were right-handed adult males (18--30 years) with full-scale IQ greater than 80, randomly selected from the ABIDE I repository and matched on age and IQ across groups. The final sample sizes (90 ASD, 92 TD) reflect the unified analysis sample obtained after quality control and Tukey IQR-based outlier removal on whole-brain coevolutionary energy.
\end{flushleft}
\end{table}

Participants were excluded if they had neurological comorbidities (such as epilepsy or traumatic brain injury), a mean framewise displacement greater than 0.5 mm or more than 20\% of volumes exceeding this criterion \cite{Faul2009}, or incomplete demographic or behavioral data. After quality control, 90 ASD and 92 TD participants remained. The two groups were matched for age (ASD: $24.3\pm3.1$ y; TD: $24.1\pm3.2$ y; $p=0.72$) and IQ (ASD: $105.2\pm9.8$; TD: $106.4\pm10.1$; $p=0.65$). All models included the scanner site as a nuisance covariate where applicable.

\subsection*{Imaging Preprocessing and fALFF Computation}
We used CPAC v1.0 \cite{Craddock:2013cpac} to preprocess all functional images. This included correcting slice timing with sinc interpolation, correcting rigid body motion with six parameters and derivatives (a total of 24 regressors), coregistering each subject's T1 image, normalizing to MNI152 (3 mm isotropic), and smoothing the images spatially with a 6 mm FWHM. In the primary analysis pipeline, nuisance regression removed linear and quadratic trends, global signal, white matter, and CSF signals, as well as motion regressors. We used a zero-phase Butterworth band-pass filter on the residual time series (0.01–0.1 Hz). We used FFT to calculate fALFF voxel-wise. ALFF was the average square-rooted power in the frequency range of 0.01 to 0.08 Hz, divided by the total power in the frequency range of 0 to 0.25 Hz \cite{Zang:2007al,Zou:2008fa}. Then, the mean fALFF was extracted for each ROI in the CC200 atlas \cite{Craddock:2012atlas}.

Because global signal regression (GSR) has been shown to introduce artifactual negative correlations that may affect the sign structure of functional connectivity \cite{Murphy2009,Fox2009,Chang2009}, and since signed link polarity is a key parameter in the coevolutionary balance framework, we additionally repeated all analyses using a preprocessing pipeline in which GSR was omitted. In this alternative pipeline, nuisance regression included only white matter and CSF signals (CompCor-based denoising; \cite{Behzadi2007}), linear and quadratic trends, and motion regressors, but did not include global signal removal. All subsequent steps (fALFF computation, parcellation, network construction, ComBat harmonization, and statistical analyses) were identical to the primary pipeline. Results from both pipelines are reported: the GSR-based analysis as the primary analysis and the no-GSR analysis as a sensitivity check.

\paragraph*{Binarization of fALFF values.}
To obtain the binary node states required by the coevolutionary Hamiltonian, each participant's regional fALFF values were thresholded at the subject-specific median. Regions with fALFF values at or above the median were assigned $s_i = +1$ (high intrinsic activity), and regions below the median were assigned $s_i = -1$ (low intrinsic activity). This median split ensures scale-free comparability across subjects and prevents bias from global fALFF amplitude differences. All coevolutionary energy and motif analyses used these binarized activity states.

\subsection*{Parcellation and Time Series Harmonization}
We used the CC200 atlas in MNI space to define 200 ROIs. To ensure consistency across all sites, all subjects' time courses were truncated to the smallest number of volumes found in any scan following preprocessing. This ensured that correlation analyses had the same number of samples.

\subsection*{Network Construction}
We built an undirected weighted graph $G=(V,E)$ with $|V|=200$ for each subject. The mean fALFF value for node $i$ was $f_i$. We computed the edge weight $w_{ij}$ as the Pearson correlation between the preprocessed time series of ROIs $i$ and $j$, followed by Fisher $z$-transformation to improve normality. No additional thresholding was applied; the complete matrix was retained for analysis.

\paragraph*{Use of full weighted functional connectivity.}
We did not threshold the functional connectivity (FC) matrices. Instead, analyses were performed on the full Fisher-$z$-transformed correlation weights, which is standard practice in modern network neuroscience \cite{Smith:2013fc}. Thresholding FC by statistical significance (e.g., $p$-values) is known to be inappropriate for resting-state data because the correlations are not independent, the temporal degrees of freedom are altered by preprocessing steps, and arbitrary thresholds can distort network topology \cite{Fornito2016}. Using the full weighted matrix preserves the graded information in both positive and negative connections and allows the Hamiltonian to capture contributions from all signed edges.

\subsection*{ComBat Harmonization}
To address multi-site batch effects inherent in pooled neuroimaging datasets, we applied ComBat harmonization \cite{Johnson2007,Fortin2017} to the derived network features. ComBat uses an empirical Bayes framework to estimate site-specific location (mean) and scale (variance) parameters and removes these batch effects while preserving biological variability of interest.

Specifically, for each network feature $y_{ij}$ (where $i$ indexes subjects and $j$ indexes features), we modeled:
\[
y_{ij} = \alpha_j + X_i \beta_j + \gamma_{i(s)j} + \delta_{i(s)j} \epsilon_{ij}
\]
where $\alpha_j$ is the overall mean for feature $j$, $X_i$ is a design matrix of biological covariates of interest (diagnostic group), $\beta_j$ is the coefficient vector, $\gamma_{i(s)j}$ is the additive site effect for subject $i$ at site $s$, $\delta_{i(s)j}$ is the multiplicative site effect, and $\epsilon_{ij}$ is the residual error. ComBat estimates these parameters using empirical Bayes shrinkage and produces harmonized data by removing the site effects.

We applied ComBat to the full feature matrix (90 features $\times$ 182 subjects) after initial quality control, using diagnostic group (ASD vs. TD) as a biological covariate to preserve. This per-feature harmonization approach ensures that site-related variance is removed independently for each coevolutionary metric while maintaining the biological signal of interest. All subsequent statistical analyses and machine learning were performed on the ComBat-harmonized features.

\subsection*{Coevolutionary Balance Theory}
We used a discrete coevolutionary balance formulation in which each brain region is assigned a binary activity state, and functional interactions contribute according to their signed connectivity weight \cite{Heider:1946ac,cartwright1956,Kargaran:2021he,GhanbarzadehNoudehi2022}. This framework extends classical structural balance theory \cite{Heider:1946ac,cartwright1956} by incorporating nodal states alongside signed links and allowing both to coevolve, yielding an energy-like quantity that reflects the degree of large-scale consistency between activity and connectivity patterns. For each participant, the nodal state $s_i \in \{-1,+1\}$ was derived by thresholding regional fALFF at the subject-specific median. For each edge $(i,j)$, the signed functional connectivity weight $w_{ij}$ was obtained from the Fisher-$z$-transformed correlation between regional BOLD time series.

The corresponding coevolutionary energy was defined as
\begin{equation}
H(G) = -\sum_{i<j} s_i \, w_{ij} \, s_j.
\label{eq:hamiltonian}
\end{equation}
This Hamiltonian reaches more negative values when regions with the same activity state are positively connected and oppositely active regions are negatively connected, and it increases when these relationships are inconsistent.

\begin{figure}[ht]
 \centering
 \includegraphics[width=0.8\linewidth]{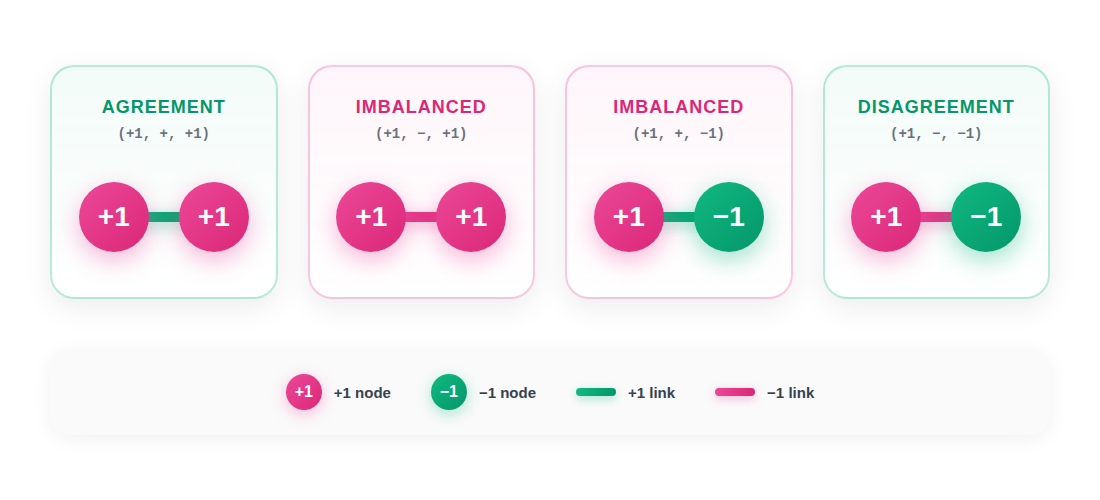}
 \caption{Pairwise configurations in the notion of coevolutionary equilibrium. The blue and orange circles show the states of the nodes ($+1$ and $-1$) representing high vs. low fALFF, while the green and red lines show the signs of the links ($+1$ and $-1$) representing positive vs. negative FC. Agreement configurations lower energy, while imbalanced pairs raise it.}
 \label{fig:configuration}
\end{figure}

\subsection*{Null-model construction}
To test whether empirical energies differed from chance expectations, we generated 1,000 topology-preserving null networks per subject. In each null replicate, nodal activity states ($s_i$) were randomly permuted across nodes, and edge signs were independently reassigned by randomly permuting the signed weights across edges, while keeping the underlying adjacency structure (i.e., the set of node pairs connected by an edge and their absolute weights) fixed. This degree-preserving randomization strategy follows the logic of the Maslov--Sneppen rewiring framework \cite{Maslov2002}, adapted here for signed weighted networks. The procedure destroys any systematic alignment between fALFF-derived nodal states and the sign structure of functional connectivity while preserving the degree distribution, weight distribution, and density of the original network \cite{Rubinov2010}. For each subject we computed the difference between the empirical energy and the mean null energy.

\subsection*{Statistical Analysis}
We utilized independent-samples $t$-tests to compare groups where the data was normal and the variance was homogeneous (Levene's test \cite{levene1960}). Otherwise, we used Mann–Whitney $U$ tests. All tests were two-tailed with $\alpha = 0.05$. Cohen's $d$ was used to report parametric effect sizes.

\paragraph*{Unified outlier removal.}
To ensure consistency across analyses, outlier removal was performed once using a single criterion based on whole-brain coevolutionary energy. For each group separately (ASD and TD), subjects whose whole-brain energy fell outside 1.5 times the interquartile range (Tukey's IQR rule) were excluded. This yielded the cleaned sample reported in Table~\ref{tab:demographics}. The same participant set was used for all subsequent analyses.

\subsection*{Connectivity, fALFF-weighted connectivity, and block-wise energy}

To relate coevolutionary energy to canonical large-scale systems, we projected the CC200 parcellation onto the seven-network cortical atlas of Yeo et al.\ \cite{Yeo2011}. For each CC200 ROI we computed the Dice overlap with each Yeo7 network and assigned the ROI to the network with maximal Dice coefficient.

For each subject we constructed a $7\times7$ block representation of connectivity and energy. For each ordered pair of networks $(u,v)$ we computed three block-wise quantities: the mean functional connectivity ($\mathrm{AvgWeight}_{uv}$), the mean fALFF-weighted connectivity ($\mathrm{AvgFalffW}_{uv}$), and the block-wise coevolutionary energy contribution ($E_{uv}$).

All connectivity and block-wise energy metrics were subjected to ComBat harmonization prior to statistical analysis, as described above.

\subsection*{Correlation Analysis with Behavioral Measures}

To assess the clinical relevance of the energy-based network measures, we conducted a comprehensive correlation analysis between all 90 ComBat-harmonized network features and eight clinical scores from the ADI-R and ADOS instruments. The clinical measures included ADI-R Social Total A, ADI-R Verbal Total BV, ADI-R RRB Total C, ADI-R Onset Total D, ADOS Total, ADOS Communication, ADOS Social, and ADOS Stereotyped Behaviors. Correlations were computed only for ASD participants with valid clinical data; subjects with missing values (coded as $-9999$ in the ABIDE phenotypic file) were excluded on a per-score basis, resulting in sample sizes ranging from $n = 36$ (ADI-R Onset Total D) to $n = 81$ (ADOS Total).

For each of the 720 feature–clinical score combinations (90 features $\times$ 8 scores), we computed both Pearson's product–moment correlation coefficient ($r$) and Spearman's rank correlation coefficient ($\rho$), along with their exact two-tailed $p$-values. To control for multiple comparisons, we applied Benjamini–Hochberg false discovery rate (FDR) correction separately to the Pearson and Spearman $p$-values across all 720 tests.

We defined nominally significant associations as those with uncorrected $p < 0.05$ for either Pearson or Spearman tests, and FDR-significant associations as those surviving correction at $q < 0.05$. All correlation analyses were performed using SciPy v1.10 and statsmodels v0.14 in Python.

\subsection*{Machine learning classification of ASD versus TD}

To examine whether the derived graph-based features carry discriminative information about diagnostic status, we performed supervised classification analyses on the adult male ABIDE I subsample with complete feature data. All 90 ComBat-harmonized features were initially considered.

Feature selection was carried out using a Random Forest classifier fitted on the training set only. We computed the mean decrease in impurity importance for each feature, ranked features by importance, and retained the top 25\% (23 features) as the final input set.

All analyses were implemented in Python using scikit-learn and XGBoost. We performed a stratified train–test split (80\%/20\%), with the held-out test set ($n = 36$) never used for model selection or tuning. Each classifier was embedded in a pipeline with feature-wise standardization estimated only on training data within each cross-validation fold.

We evaluated five classifiers with the following hyperparameter grids:
\begin{itemize}
    \item \textbf{SVM (RBF)}: $C \in \{0.1, 1, 10, 100\}$, $\gamma \in \{\text{scale}, 0.001, 0.0001\}$
    \item \textbf{XGBoost}: max\_depth $\in \{2, 3, 5\}$, n\_estimators $\in \{100, 200, 300\}$, learning\_rate $\in \{0.01, 0.1, 0.2\}$, subsample $\in \{0.7, 0.8, 1.0\}$, colsample\_bytree $\in \{0.7, 0.8, 1.0\}$
    \item \textbf{Logistic Regression}: $C \in \{0.01, 0.1, 0.5, 1, 10\}$, penalty $\in \{$L1, L2$\}$, solver $\in \{$liblinear$\}$
    \item \textbf{Gaussian Naive Bayes}: var\_smoothing $\in \{10^{-9}, 10^{-8}, 10^{-7}\}$
    \item \textbf{KNN}: n\_neighbors $\in \{3, 5, 7, 9, 11, 13, 14, 15\}$, metric $\in \{$euclidean, manhattan$\}$
\end{itemize}

Grid search with stratified 5-fold cross-validation selected the best hyperparameters for each model based on mean accuracy. The best configuration was refitted on the entire training set and evaluated once on the held-out test set. Confidence intervals were computed via bootstrap resampling (1000 iterations).

\section*{Data availability}

This study used only de-identified resting-state fMRI and phenotypic data from the Autism Brain Imaging Data Exchange I (ABIDE I) repository \cite{DiMartino:2014abide}. All raw imaging and behavioral data are publicly available from the ABIDE initiative (http://fcon\_1000.projects.nitrc.org/indi/abide/). The derived coevolutionary energy metrics and graph-based feature matrices generated during the current study are available from the corresponding author on reasonable request.

\section*{Code availability}

All custom code used to compute coevolutionary energy, motif statistics, ComBat harmonization, and machine-learning models was written in Python using standard open-source libraries (NumPy, SciPy, scikit-learn, XGBoost, and neuroCombat). The scripts used for preprocessing of graph features and for reproducing the main analyses are available from the corresponding author on reasonable request.

\section*{Ethics statement}

All imaging and behavioral data analyzed in this study were obtained from the publicly available ABIDE I repository \cite{DiMartino:2014abide}. Data collection at each contributing site was approved by the local institutional review board, and written informed consent was obtained from all participants or their legal guardians, in accordance with the Declaration of Helsinki. The present analyses were conducted on fully de-identified data and did not involve any direct interaction with participants; as such, no additional ethical approval was required at the authors' institution.

\section*{Competing interests}

The authors declare no competing interests.

\section*{Author contributions}

S. Rezaei Afshar conceived the study. G.R. Jafari designed the analysis pipeline. S. Rezaei Afshar implemented the coevolutionary energy framework, applied ComBat harmonization, and performed the statistical analyses. G.R. Jafari contributed to interpretation of the results and neurobiological framing. S. Rezaei Afshar drafted the manuscript, and all authors revised the manuscript critically for important intellectual content. All authors approved the final version of the manuscript.

\bibliography{Bibliography}

@ARTICLE{Kargaran:2021he,
   author = {Kargaran, A. and Jafari, G.~R.},
   title = {Heider and coevolutionary balance: From discrete to continuous phase transition},
   journal = {Physical Review E},
   volume = {103},
   pages = {052302},
   year = {2021},
   doi = {https://doi.org/10.1103/PhysRevE.103.052302}
}

@BOOK{APA:2022ds,
   author = {{American Psychiatric Association}},
   title = {Diagnostic and Statistical Manual of Mental Disorders, Fifth Edition, Text Revision (DSM-5-TR)},
   publisher = {American Psychiatric Association Publishing},
   year = {2022},
   isbn = {978-0890425756},
   address = {Washington, DC}
}

@ARTICLE{Zang:2007al,
   author = {Zang, Yu-Feng et al.},
   title = {Altered baseline brain activity in children with ADHD revealed by resting-state functional MRI},
   journal = {Brain and Development},
   year = {2007},
   doi = {10.1016/j.braindev.2006.07.002}
}

@ARTICLE{Heider:1946ac,
   author = {Heider, Fritz},
   title = {Attitudes and Cognitive Organization},
   journal = {The Journal of Psychology: Interdisciplinary and Applied},
   volume = {21},
   number = {1},
   pages = {107–112},
   year = {1946},
   doi = {10.1080/00223980.1946.9917275}
}

@ARTICLE{Maenner:2023pr,
   author = {Maenner, Matthew J. et al.},
   title = {Prevalence and Characteristics of Autism Spectrum Disorder Among Children Aged 8 Years — Autism and Developmental Disabilities Monitoring Network, 11 Sites, United States, 2020},
   journal = {Morbidity and Mortality Weekly Report: Surveillance Summaries},
   volume = {72},
   number = {2},
   year = {2023}
}

@ARTICLE{Zou:2008fa,
   author = {Zou, Qi-Hong et al.},
   title = {An improved approach to detection of amplitude of low-frequency fluctuation (ALFF) for resting-state fMRI: Fractional ALFF},
   journal = {Journal of Neuroscience Methods},
   volume = {172},
   number = {1},
   pages = {137–141},
   year = {2008},
   doi = {10.1016/j.jneumeth.2008.04.012}
}

@ARTICLE{Smith:2013fc,
   author = {Smith, Stephen M. et al.},
   title = {Functional connectomics from resting-state fMRI},
   journal = {Trends in Cognitive Sciences},
   volume = {17},
   number = {12},
   pages = {666–682},
   year = {2013},
   doi = {10.1016/j.tics.2013.09.016}
}

@ARTICLE{DiMartino:2014abide,
   author = {Di Martino, Adriana and Yan, Chao-Gan and Li, Qingyang and Denio, Erin and Castellanos, Francisco X. and Alaerts, Kaat and Anderson, Jeffrey S. and Assaf, Michal and Bookheimer, Susan Y. and Dapretto, Mirella and Deen, Ben and Delmonte, Sonja and Dinstein, Ilan and Ertl-Wagner, Birgit and Fair, Damien A. and Gallagher, Louise and Kennedy, Daniel P. and Keown, Christopher L. and Keysers, Christian and Lainhart, Janet E. and Lord, Catherine and Luna, Beatriz and Menon, Vinod and Minshew, Nancy and Monk, Christopher S. and Mueller, Sophia and Müller, Ralph-Axel and Nebel, Mary Beth and Nigg, Joel T. and O’Hearn, Kirsten and Pelphrey, Kevin A. and Peltier, Scott J. and Rudie, Jeffrey D. and Sunaert, Stefan and Thioux, Marc and Tyszka, J. Michael and Uddin, Lucina Q. and Verhoeven, Judith S. and Wenderoth, Nicole and Wiggins, Jillian L. and Mostofsky, Stewart H. and Milham, Michael P.},
   title = {The Autism Brain Imaging Data Exchange: Towards a Large-Scale Evaluation of the Intrinsic Brain Architecture in Autism},
   journal = {Molecular Psychiatry},
   volume = {19},
   number = {6},
   pages = {659--667},
   year = {2014},
   doi = {10.1038/mp.2013.78}
}

@ARTICLE{Craddock:2013cpac,
  author = {Craddock, Cameron R. and Sikka, Satrajit S. and Cheung, Brian and Khanuja, Rohan and Ghosh, Suman and Yan, Chao-Gan and Li, Qingyang and Lurie, Daniel and Vogelstein, Joshua and Castellanos, Francisco X. and Di Martino, Adriana and Milham, Michael P.},
  title = {Towards automated analysis of connectomes: The Configurable Pipeline for the Analysis of Connectomes (C-PAC)},
  journal = {Frontiers in Neuroinformatics},
  volume = {7},
  pages = {42},
  year = {2013},
  doi = {10.3389/fninf.2013.00042}
}

@ARTICLE{Craddock:2012atlas,
  author = {Craddock, Cameron R. and James, G. Andrew and Holtzheimer, Paul E. and Hu, Xiaoping P. and Mayberg, Helen S.},
  title = {A whole brain fMRI atlas generated via spatially constrained spectral clustering},
  journal = {Human Brain Mapping},
  volume = {33},
  number = {8},
  pages = {1914--1928},
  year = {2012},
  doi = {10.1002/hbm.21333}
}

@article{cartwright1956,
  author = {Cartwright, Dorwin and Harary, Frank},
  title = {Structural balance: A generalization of Heider's theory},
  journal = {Psychological Review},
  year = {1956},
  volume = {63},
  pages = {277--293}
}

@article{levene1960,
  author = {Levene, Howard},
  title = {Robust Tests for Equality of Variances},
  journal = {Contributions to Probability and Statistics: Essays in Honor of Harold Hotelling},
  year = {1960},
  publisher = {Stanford University Press},
  pages = {278--292}
}

@article{Yeo2011,
  author    = {Yeo, B. T. Thomas and Krienen, Fenna M. and Sepulcre, Jorge and Sabuncu, Mert R. and Lashkari, Danial and Hollinshead, Marisa and Roffman, Joshua L. and Smoller, Jordan W. and Z{\"o}llei, Lilla and Polimeni, Jonathan R. and Fischl, Bruce and Liu, Hesheng and Buckner, Randy L.},
  title     = {The organization of the human cerebral cortex estimated by intrinsic functional connectivity},
  journal   = {Journal of Neurophysiology},
  volume    = {106},
  number    = {3},
  pages     = {1125--1165},
  year      = {2011},
  doi       = {10.1152/jn.00338.2011},
  url       = {https://doi.org/10.1152/jn.00338.2011}
}

@article{uddin2013,
  author = {Uddin, Lucina Q. and Supekar, Kaustubh and Lynch, Christopher J. and Khouzam, Anjali and Phillips, Jennifer and Feinstein, Caroline and Ryali, Srikanth and Menon, Vinod},
  title = {Salience network–based classification and prediction of symptom severity in children with autism},
  journal = {JAMA Psychiatry},
  volume = {70},
  number = {8},
  pages = {869--879},
  year = {2013},
  doi = {10.1001/jamapsychiatry.2013.104}
}

@ARTICLE{10.3389/fnhum.2013.00458,
AUTHOR={Uddin, Lucina Q. and Supekar, Kaustubh  and Menon, Vinod },
TITLE={Reconceptualizing functional brain connectivity in autism from a developmental perspective},      
JOURNAL={Frontiers in Human Neuroscience},     
VOLUME={Volume 7 - 2013},
YEAR={2013},
URL={https://www.frontiersin.org/journals/human-neuroscience/articles/10.3389/fnhum.2013.00458},
DOI={10.3389/fnhum.2013.00458},
ISSN={1662-5161}}

@article{Faul2009,
  author = {Faul, Franz and Erdfelder, Edgar and Buchner, Axel and Lang, Albert-Georg},
  title = {Statistical power analyses using G*Power 3.1: Tests for correlation and regression analyses},
  journal = {Behavior Research Methods},
  year = {2009},
  volume = {41},
  number = {4},
  pages = {1149--1160},
  doi = {10.3758/BRM.41.4.1149}
}

@article{scikit-learn,
  title        = {Scikit-learn: Machine Learning in {P}ython},
  author       = {Pedregosa, F. and Varoquaux, G. and Gramfort, A. and Michel, V.
                  and Thirion, B. and Grisel, O. and Blondel, M. and Prettenhofer, P.
                  and Weiss, R. and Dubourg, V. and Vanderplas, J. and Passos, A. and
                  Cournapeau, D. and Brucher, M. and Perrot, M. and Duchesnay, E.},
  journal      = {Journal of Machine Learning Research},
  volume       = {12},
  pages        = {2825--2830},
  year         = {2011},
  url          = {http://www.jmlr.org/papers/volume12/pedregosa11a/pedregosa11a.pdf}
}

@article{Rubinov2010,
  title={Complex network measures of brain connectivity: Uses and interpretations},
  author={Rubinov, Mikail and Sporns, Olaf},
  journal={NeuroImage},
  volume={52},
  number={3},
  pages={1059--1069},
  year={2010}
}

@article{assaf2010abnormal,
  author       = {Assaf, M. and Jagannathan, K. and Calhoun, V.~D. and Miller, L. and Stevens, M.~C. and Sahl, R. and Pearlson, G.~D. and Adalı, T.},
  title        = {Abnormal functional connectivity of default mode sub‐networks in autism spectrum disorder patients},
  journal      = {NeuroImage},
  volume       = {53},
  pages        = {247--256},
  year         = {2010},
  doi          = {10.1016/j.neuroimage.2010.05.067},
}

@article{GhanbarzadehNoudehi2022,
  author  = {Ghanbarzadeh Noudehi, M. and Kargaran, A. and Azimi-Tafreshi, N. and Jafari, G.~R.},
  title   = {Second- to first-order phase transition: Coevolutionary versus structural balance},
  journal = {Phys. Rev. E},
  volume  = {106},
  pages   = {044303},
  year    = {2022},
  doi     = {10.1103/PhysRevE.106.044303},
}

@article{Yizhar2011,
  author    = {Yizhar, Ofer and Fenno, Laurie E. and Prigge, Madeleine and Schneider, Fabian and Davidson, Thomas J. and O’Shea, Deneen J. and Sohal, Valentina S. and Goshen, Inbal and Finkelstein, Joel and Paz, Joseph T. and Stehfest, Karen and Fudim, Rachel and Ramakrishnan, Chidambaram and Huguenard, John R. and Hegemann, Peter and Deisseroth, Karl},
  title     = {Neocortical excitation/inhibition balance in information processing and social dysfunction},
  journal   = {Nature},
  volume    = {477},
  pages     = {171--178},
  year      = {2011},
  doi       = {10.1038/nature10360},
}

@article{Dinstein2011,
  author    = {Dinstein, Ilan and Pierce, Karen and Eyler, Lisa and Solso, Samuel and Malach, Rafael and Behrmann, Marlene and Courchesne, Eric},
  title     = {Disrupted neural synchronization in toddlers with autism},
  journal   = {Neuron},
  volume    = {70},
  number    = {6},
  pages     = {1218--1225},
  year      = {2011},
  doi       = {10.1016/j.neuron.2011.05.018},
}

@book{Fornito2016,
  title={Fundamentals of Brain Network Analysis},
  author={Fornito, Alex and Zalesky, Andrew and Bullmore, Edward},
  year={2016},
  publisher={Academic Press}
}

@article{Fortin2017,
  author    = {Fortin, Jean-Philippe and Parker, Drew and Tun{\c{c}}, Birkan and Watanabe, Takanori and Elliott, Mark A. and Ruparel, Kosha and Roalf, David R. and Satterthwaite, Theodore D. and Gur, Ruben C. and Gur, Raquel E. and Schultz, Robert T. and Verma, Ragini and Shinohara, Russell T.},
  title     = {Harmonization of multi-site diffusion tensor imaging data},
  journal   = {NeuroImage},
  volume    = {161},
  pages     = {149--170},
  year      = {2017},
  doi       = {10.1016/j.neuroimage.2017.08.047},
  pmid      = {28826946}
}

@article{Johnson2007,
  author    = {Johnson, W. Evan and Li, Cheng and Rabinovic, Ariel},
  title     = {Adjusting batch effects in microarray expression data using empirical {B}ayes methods},
  journal   = {Biostatistics},
  volume    = {8},
  number    = {1},
  pages     = {118--127},
  year      = {2007},
  doi       = {10.1093/biostatistics/kxj037},
  pmid      = {16632515}
}

@article{Maslov2002,
  author  = {Maslov, Sergei and Sneppen, Kim},
  title   = {Specificity and stability in topology of protein networks},
  journal = {Science},
  volume  = {296},
  number  = {5569},
  pages   = {910--913},
  year    = {2002},
  doi     = {10.1126/science.1065103}
}

@article{Murphy2009,
  author  = {Murphy, Kevin and Birn, Rasmus M. and Handwerker, Daniel A. and Jones, Tyler B. and Bandettini, Peter A.},
  title   = {The impact of global signal regression on resting state correlations: Are anti-correlated networks introduced?},
  journal = {NeuroImage},
  volume  = {44},
  number  = {3},
  pages   = {893--905},
  year    = {2009},
  doi     = {10.1016/j.neuroimage.2008.09.036}
}

@article{Fox2009,
  author  = {Fox, Michael D. and Zhang, Dongyang and Snyder, Abraham Z. and Raichle, Marcus E.},
  title   = {The global signal and observed anticorrelated resting state brain networks},
  journal = {Journal of Neurophysiology},
  volume  = {101},
  number  = {6},
  pages   = {3270--3283},
  year    = {2009},
  doi     = {10.1152/jn.90777.2008}
}

@article{Chang2009,
  author  = {Chang, Catie and Glover, Gary H.},
  title   = {Effects of model-based physiological noise correction on default mode network anti-correlations and correlations},
  journal = {NeuroImage},
  volume  = {47},
  number  = {4},
  pages   = {1448--1459},
  year    = {2009},
  doi     = {10.1016/j.neuroimage.2009.05.012}
}

@article{Behzadi2007,
  author  = {Behzadi, Yashar and Restom, Khaled and Liau, Joy and Liu, Thomas T.},
  title   = {A component based noise correction method ({CompCor}) for {BOLD} and perfusion based {fMRI}},
  journal = {NeuroImage},
  volume  = {37},
  number  = {1},
  pages   = {90--101},
  year    = {2007},
  doi     = {10.1016/j.neuroimage.2007.04.042}
}

\section*{Funding}

This project did not receive any outside funding.

\clearpage
\section*{Supplementary Material}
\renewcommand{\thetable}{S\arabic{table}}

\begin{table}[htbp]
\centering
\caption{\textbf{Stratified group comparisons for all whole-brain coevolutionary metrics after ComBat harmonization.} 
For each feature, the table reports group means and standard deviations, the test used, raw two-tailed $p$-value, FDR-corrected $p$-value, and significance status. All tests used the unified outlier-cleaned sample (90 ASD, 92 TD).}
\label{tab:supp_whole_stats}
\scriptsize
\begin{tabular}{lcccccccc}
\toprule
\textbf{Feature} & \textbf{ASD Mean} & \textbf{ASD SD} & \textbf{TD Mean} & \textbf{TD SD} & \textbf{Test} & \textbf{$p$} & \textbf{$p_{\mathrm{FDR}}$} & \textbf{Sig.} \\
\midrule
Energy\_global 
  & $-264.92$ & $110.28$ 
  & $-210.85$ & $83.51$ 
  & Mann–Whitney 
  & $8.35\times 10^{-4}$ & $1.95\times 10^{-3}$ & Yes \\
Prop\_agreement 
  & $0.2502$ & $0.0063$ 
  & $0.2475$ & $0.0045$ 
  & Mann–Whitney 
  & $9.74\times 10^{-4}$ & $1.95\times 10^{-3}$ & Yes \\
Prop\_imbalanced\_same 
  & $0.2473$ & $0.0063$ 
  & $0.2500$ & $0.0044$ 
  & Mann–Whitney 
  & $9.25\times 10^{-4}$ & $1.95\times 10^{-3}$ & Yes \\
Prop\_disagreement 
  & $0.2695$ & $0.0057$ 
  & $0.2673$ & $0.0052$ 
  & $t$-test 
  & $7.79\times 10^{-3}$ & $1.00\times 10^{-2}$ & Yes \\
Prop\_imbalanced\_opp 
  & $0.2330$ & $0.0057$ 
  & $0.2352$ & $0.0053$ 
  & $t$-test 
  & $8.35\times 10^{-3}$ & $1.00\times 10^{-2}$ & Yes \\
Bipolarity 
  & $0.6173$ & $0.0259$ 
  & $0.6174$ & $0.0313$ 
  & Mann–Whitney 
  & $0.419$ & $0.419$ & No \\
\bottomrule
\end{tabular}
\end{table}

\begin{table}[htbp]
\centering
\caption{\textbf{Full random forest feature importance ranking.}
All 90 features ranked by mean decrease in impurity from the Random Forest classifier fitted on training data. The top 23 features (top 25\%) were retained for classification.}
\label{tab:selected_features}
\scriptsize
\begin{tabular}{rlc|rlc}
\toprule
\textbf{Rank} & \textbf{Feature} & \textbf{Imp.} & \textbf{Rank} & \textbf{Feature} & \textbf{Imp.} \\
\midrule
1 & Inter\_Vis\_\_Sal/VA\_AvgWt & 0.025 & 24 & Inter\_Sal/VA\_\_FP\_AvgWt & 0.014 \\
2 & Inter\_Vis\_\_SM\_AvgWt & 0.024 & 25 & Inter\_FP\_\_Def\_AvgWt & 0.013 \\
3 & Whole\_Prop\_imb\_same & 0.023 & 26 & Inter\_DA\_\_Sal/VA\_EnBl & 0.013 \\
4 & Whole\_Prop\_agreement & 0.021 & 27 & Inter\_DA\_\_Sal/VA\_AvgFW & 0.013 \\
5 & Intra\_DA\_AvgWt & 0.020 & 28 & Inter\_SM\_\_Sal/VA\_AvgWt & 0.013 \\
6 & Inter\_SM\_\_Lim\_AvgWt & 0.019 & 29 & Intra\_SM\_AvgWt & 0.012 \\
7 & Inter\_Lim\_\_Def\_EnBl & 0.019 & 30 & Inter\_Vis\_\_Lim\_EnBl & 0.012 \\
8 & Inter\_DA\_\_Def\_AvgWt & 0.018 & \multicolumn{3}{c}{...} \\
9 & Intra\_Lim\_EnBl & 0.018 & 88 & Inter\_SM\_\_DA\_EnBl & 0.005 \\
10 & Intra\_Vis\_EnBl & 0.017 & 89 & Inter\_Sal/VA\_\_Def\_AvgFW & 0.005 \\
11 & Inter\_DA\_\_Lim\_AvgFW & 0.016 & 90 & Intra\_Def\_AvgWt & 0.004 \\
\bottomrule
\end{tabular}
\begin{flushleft}
\footnotesize
Abbreviations: Vis=Visual, SM=SomatoMotor, DA=DorsalAttn, Sal/VA=Salience/VentAttn, Lim=Limbic, FP=Frontoparietal, Def=Default, AvgWt=AvgWeight, AvgFW=AvgFalffW, EnBl=EnergyBlock, imb=imbalanced.
\end{flushleft}
\end{table}

\begin{table}[htbp]
\centering
\caption{\textbf{Performance of supervised classifiers for ASD versus TD classification with best hyperparameters.}
Each row corresponds to the best configuration identified by grid search. CV score is mean accuracy from stratified 5-fold cross-validation on training data. Test metrics were computed on the held-out test set ($n = 36$). 95\% CIs were computed via bootstrap.}
\label{tab:supp_ml_results}
\scriptsize
\begin{tabular}{p{0.15\linewidth}p{0.42\linewidth}cccc}
\toprule
\textbf{Model} & \textbf{Best Hyperparameters} & \textbf{CV} & \textbf{Test Acc} & \textbf{Test F1} & \textbf{AUC} \\
\midrule
GaussianNB 
  & var\_smoothing $= 10^{-9}$ 
  & 0.665 & 0.778 & 0.771 & 0.79 \\
SVM (RBF)
  & $C = 100$, $\gamma = 0.0001$ 
  & 0.672 & 0.694 & 0.692 & ,  \\
KNN 
  & $k = 14$, metric = euclidean 
  & 0.653 & 0.694 & 0.674 & ,  \\
XGBoost 
  & max\_depth = 2, n\_est = 300, lr = 0.2, subsample = 0.7, colsample = 0.7 
  & 0.666 & 0.611 & 0.600 & ,  \\
LogReg 
  & $C = 0.5$, penalty = L2, solver = liblinear 
  & 0.665 & 0.583 & 0.567 & ,  \\
\bottomrule
\end{tabular}
\end{table}


\renewcommand{\thefigure}{S\arabic{figure}}
\setcounter{figure}{0}

\begin{figure}[htbp]
  \centering
  \includegraphics[width=0.95\linewidth]{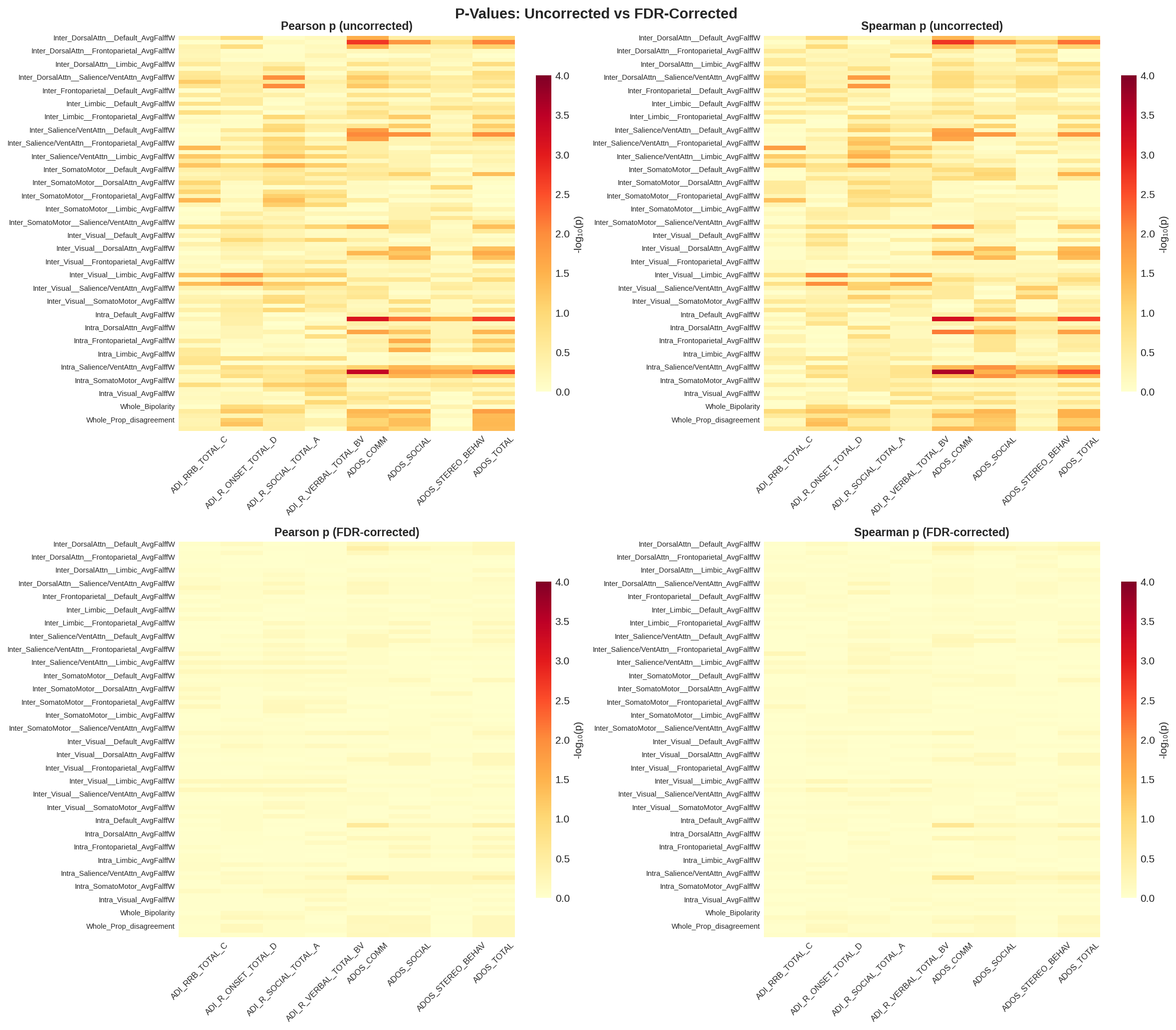}
  \caption{\textbf{P-value heatmaps for feature–clinical score correlations.} Top row: uncorrected $p$-values for Pearson (left) and Spearman (right) correlations. Bottom row: FDR-corrected $p$-values. Color scale shows $-\log_{10}(p)$, with warmer colors indicating smaller $p$-values. None of the 720 tests survived FDR correction at $q < 0.05$.}
  \label{fig:supp_pval_heatmap}
\end{figure}

\begin{figure}[htbp]
  \centering
  \includegraphics[width=0.7\linewidth]{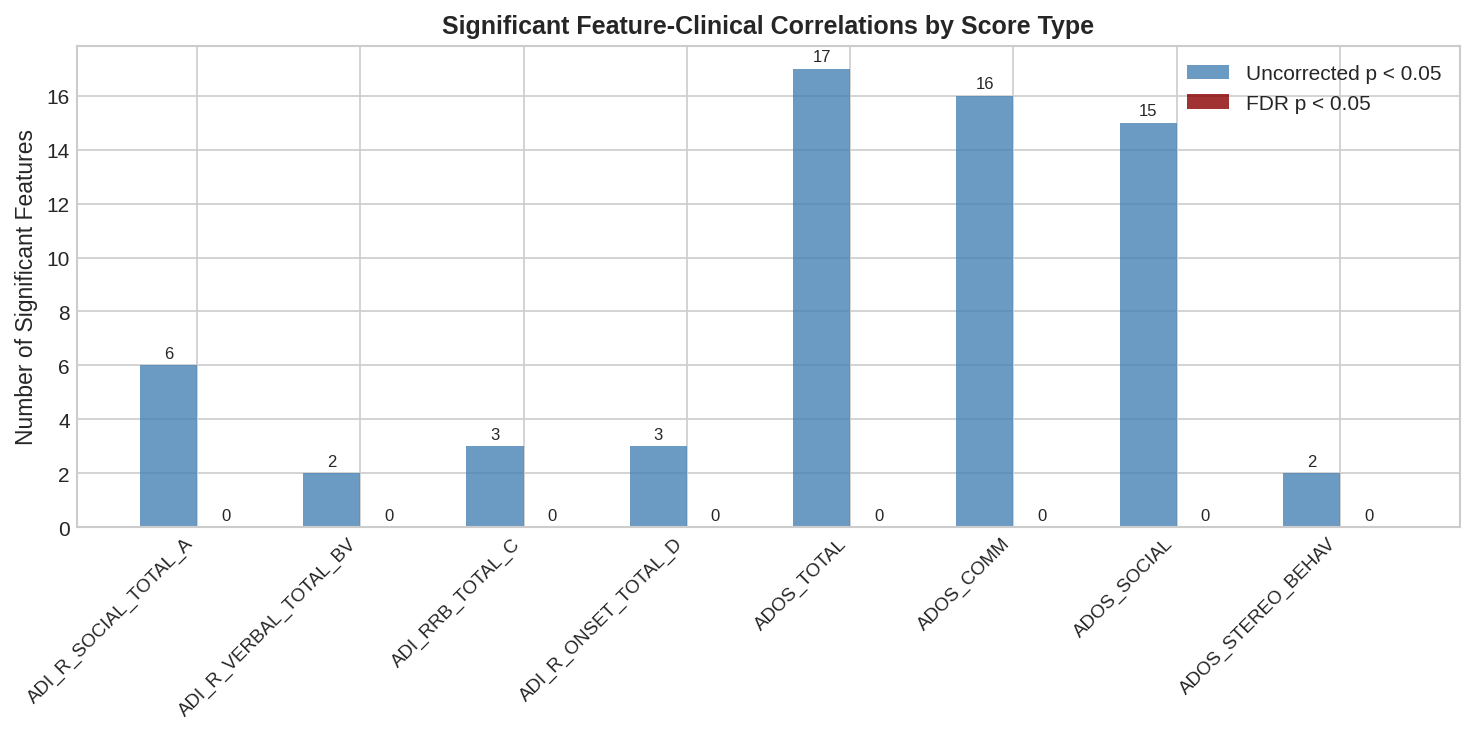}
  \caption{\textbf{Summary of significant correlations by clinical score.} Bar chart showing the number of network features with nominally significant correlations (uncorrected $p < 0.05$) for each of the eight clinical scores. Blue bars: uncorrected significance. Red bars: FDR-corrected significance (all zero). ADOS measures showed more significant associations than ADI-R subscales, likely reflecting larger sample sizes for ADOS ($n = 63$--81) compared to ADI-R ($n = 36$--39).}
  \label{fig:supp_sig_summary}
\end{figure}

\begin{table}[htbp]
\centering
\caption{\textbf{Full list of nominally significant feature–clinical score correlations (uncorrected $p < 0.05$).} 
The comprehensive analysis tested 720 correlations (90 features $\times$ 8 clinical scores). This table lists all 64 pairs that reached uncorrected significance for either Pearson or Spearman tests. None survived FDR correction. Sample sizes vary due to missing clinical data.}
\label{tab:supp_corr_sig}
\scriptsize
\begin{tabular}{p{4.5cm}lccccc}
\toprule
\textbf{Feature} & \textbf{Clinical Score} & \textbf{$n$} & \textbf{$r$} & \textbf{$p$} & \textbf{$\rho$} & \textbf{$p^{\rho}$} \\
\midrule
Intra\_Salience/VentAttn\_AvgWeight & ADOS\_COMM & 71 & $-$0.41 & 0.0004 & $-$0.42 & 0.0002 \\
Intra\_Default\_AvgWeight & ADOS\_COMM & 71 & $-$0.39 & 0.0007 & $-$0.40 & 0.0006 \\
Inter\_DorsalAttn\_\_Default\_AvgWeight & ADOS\_COMM & 71 & 0.36 & 0.0019 & 0.36 & 0.0018 \\
Intra\_Default\_AvgWeight & ADOS\_TOTAL & 81 & $-$0.34 & 0.0022 & $-$0.33 & 0.0027 \\
Intra\_Salience/VentAttn\_AvgWeight & ADOS\_TOTAL & 81 & $-$0.33 & 0.0029 & $-$0.32 & 0.0032 \\
Intra\_Default\_AvgWeight & ADOS\_SOCIAL & 71 & $-$0.32 & 0.0072 & $-$0.30 & 0.0104 \\
Inter\_DorsalAttn\_\_Default\_AvgWeight & ADOS\_TOTAL & 81 & 0.29 & 0.0084 & 0.30 & 0.0061 \\
Inter\_Salience/VentAttn\_\_Default\_AvgWeight & ADOS\_COMM & 71 & 0.31 & 0.0094 & 0.28 & 0.0167 \\
Inter\_DorsalAttn\_\_Salience/VentAttn\_EnergyBlock & ADI\_R\_SOCIAL\_TOTAL\_A & 39 & $-$0.41 & 0.0099 & $-$0.39 & 0.0147 \\
Inter\_DorsalAttn\_\_Salience/VentAttn\_AvgFalffW & ADI\_R\_SOCIAL\_TOTAL\_A & 39 & 0.41 & 0.0104 & 0.38 & 0.0159 \\
Inter\_Salience/VentAttn\_\_Default\_AvgWeight & ADOS\_TOTAL & 81 & 0.28 & 0.0104 & 0.28 & 0.0129 \\
Inter\_Salience/VentAttn\_\_Default\_AvgWeight & ADOS\_SOCIAL & 71 & 0.30 & 0.0110 & 0.29 & 0.0154 \\
Inter\_DorsalAttn\_\_Default\_AvgWeight & ADOS\_SOCIAL & 71 & 0.29 & 0.0131 & 0.30 & 0.0104 \\
Inter\_Visual\_\_Limbic\_AvgFalffW & ADI\_R\_ONSET\_TOTAL\_D & 36 & 0.40 & 0.0155 & 0.43 & 0.0092 \\
Inter\_Salience/VentAttn\_\_Default\_AvgFalffW & ADOS\_COMM & 71 & 0.28 & 0.0165 & 0.27 & 0.0208 \\
Inter\_Salience/VentAttn\_\_Default\_EnergyBlock & ADOS\_COMM & 71 & $-$0.28 & 0.0168 & $-$0.27 & 0.0218 \\
Whole\_Energy\_global & ADOS\_TOTAL & 81 & $-$0.26 & 0.0169 & $-$0.24 & 0.0306 \\
Inter\_Visual\_\_Limbic\_EnergyBlock & ADI\_R\_ONSET\_TOTAL\_D & 36 & $-$0.39 & 0.0175 & $-$0.42 & 0.0101 \\
Intra\_Salience/VentAttn\_AvgWeight & ADOS\_SOCIAL & 71 & $-$0.28 & 0.0202 & $-$0.26 & 0.0318 \\
Intra\_DorsalAttn\_AvgWeight & ADOS\_COMM & 71 & $-$0.27 & 0.0209 & $-$0.32 & 0.0073 \\
\multicolumn{7}{c}{\textit{... 44 additional rows (see online supplementary data) ...}} \\
\bottomrule
\end{tabular}
\begin{flushleft}
\footnotesize
Note: The full table with all 64 nominally significant correlations is available in the online supplementary data file \texttt{significant\_uncorrected.csv}. FDR-corrected $p$-values ranged from 0.17 to 0.87; none reached significance at $q < 0.05$.
\end{flushleft}
\end{table}

\begin{table}[htbp]
\centering
\caption{\textbf{Summary statistics for the comprehensive correlation analysis.} 
Overview of the feature–clinical score correlation analysis, including sample sizes, number of tests, and significance thresholds.}
\label{tab:supp_corr_full}
\small
\begin{tabular}{lc}
\toprule
\textbf{Parameter} & \textbf{Value} \\
\midrule
Total network features tested & 90 \\
Clinical scores tested & 8 \\
Total correlation tests & 720 \\
Subjects with matched data & 85 \\
Sample size range (per clinical score) & 36--81 \\
\midrule
\multicolumn{2}{l}{\textit{Clinical scores and sample sizes:}} \\
\quad ADOS\_TOTAL & $n = 81$ \\
\quad ADOS\_COMM & $n = 71$ \\
\quad ADOS\_SOCIAL & $n = 71$ \\
\quad ADOS\_STEREO\_BEHAV & $n = 63$ \\
\quad ADI\_R\_SOCIAL\_TOTAL\_A & $n = 39$ \\
\quad ADI\_R\_VERBAL\_TOTAL\_BV & $n = 39$ \\
\quad ADI\_RRB\_TOTAL\_C & $n = 39$ \\
\quad ADI\_R\_ONSET\_TOTAL\_D & $n = 36$ \\
\midrule
Nominally significant (uncorrected $p < 0.05$) & 64 (8.9\%) \\
FDR-significant ($q < 0.05$) & 0 (0\%) \\
Strongest Pearson $r$ & $-$0.41 \\
Smallest uncorrected $p$ & 0.0004 \\
\bottomrule
\end{tabular}
\end{table}

\end{document}